\documentclass[fleqn,usenatbib]{mnras}

\usepackage{newtxtext,newtxmath}
\usepackage{xcolor}
\usepackage[T1]{fontenc}
\usepackage{lipsum}

\usepackage{ae,aecompl}

\usepackage{graphicx}	
\usepackage{amsmath}	
\usepackage{amssymb}	
\usepackage{multirow}
\usepackage{rotating}
\usepackage{array,makecell}
\usepackage{longtable}
\usepackage[lined]{algorithm2e}

\DeclareMathOperator*{\argmin}{argmin}
\DeclareMathOperator*{\argmax}{argmax}
\SetAlCapSkip{1em}

\title[Active Learning for SN photometric classification]{Optimizing spectroscopic follow-up strategies for supernova photometric classification with active learning} 

\author[Ishida \textit{et al.}]{E. E. O. Ishida,$^{1}$\thanks{E-mail:  emille.ishida@clermont.in2p3.fr (EEOI)}
R. Beck,$^{2,3}$,
S. Gonz\'alez-Gait\'an$^{4}$,
R. S. de Souza$^{5}$,\newauthor
A. Krone-Martins$^{6}$,
J. W. Barrett$^{7}$,
N. Kennamer$^{8}$,
R. Vilalta$^{9}$,
J. M. Burgess$^{10}$,\newauthor
B. Quint$^{11}$,
A. Z. Vitorelli$^{12}$,
A. Mahabal$^{13}$, 
and E. Gangler$^{1}$, \newauthor
for the COIN collaboration
\\
$^{1}$Universit\'e Clermont Auvergne, CNRS/IN2P3, LPC, F-63000 Clermont-Ferrand, France\\
$^{2}$Institute for Astronomy, University of Hawaii, 2680 Woodlawn Drive, Honolulu, HI, 96822, USA\\
$^{3}$Department of Physics of Complex Systems, E\"{o}tv\"{o}s Lor\'and University, Pf. 32, H-1518 Budapest, Hungary\\
$^{4}$CENTRA/COSTAR, Instituto Superior T\'ecnico, Universidade de Lisboa, Av. Rovisco Pais 1, 1049-001 Lisboa, Portugal\\
$^{5}$Department of Physics \& Astronomy, University of North Carolina at Chapel Hill, Chapel Hill, NC 27599-3255, USA\\
$^{6}$CENTRA/SIM, Faculdade de Ci\^encias, Universidade de Lisboa, Ed. C8, Campo Grande, 1749-016, Lisboa, Portugal\\
$^{7}$Institute of Gravitational-wave Astronomy and School of Physics and Astronomy, University of Birmingham, Edgbaston, \\ Birmingham, B15 2TT, United Kingdom\\
$^{8}$Department of Computer Science, University of California Irvine, Donald Bren Hall, Irvine, CA 92617, USA\\
$^{9}$Department of Computer Science, University of Houston, 
3551 Cullen Blvd, 501 PGH, Houston, Texas 77204-3010, USA\\
$^{10}$Max-Planck-Institut f\"ur extraterrestrische Physik, Giessenbachstrasse, D-85748 Garching, Germany\\
$^{11}$SOAR Telescope, AURA-O, Colina El Pino S/N, Casila 603, La Serena, Chile\\
$^{12}$Instituto de Astronomia, Geof\'isica e Ci\^encias Atmosf\'ericas, Universidade de S\~ao Paulo, S\~ao Paulo, SP, Brazil\\
$^{13}$Center for Data-Driven Discovery, California Institute of Technology, Pasadena, CA, 91125, USA
}

\date{Accepted XXX. Received YYY; in original form ZZZ}

\pubyear{2018}

\begin{document}
\label{firstpage}
\pagerange{\pageref{firstpage}--\pageref{lastpage}}
\maketitle

\begin{abstract}
We report a framework for spectroscopic follow-up design for optimizing supernova photometric classification. The strategy accounts for the unavoidable mismatch between spectroscopic and photometric samples, and can be used even in the beginning of a new survey -  without  any initial training set. The framework falls under the umbrella of active learning (AL), a class of algorithms that aims to minimize labelling costs by identifying  a few, carefully chosen, objects which have high potential in improving the classifier predictions. As a proof of concept, we use the simulated data released after the Supernova Photometric Classification Challenge (SNPCC) and a random forest classifier. Our results show that,  using only 12\% the number of training objects in the SNPCC spectroscopic sample, this approach is able to double purity results. 
Moreover, in order to take into account multiple spectroscopic observations in the same night, we propose a semi-supervised batch-mode AL algorithm which selects a set of $N=5$ most informative objects at each night. In comparison with the initial state using the traditional approach, our method achieves 2.3 times higher purity and comparable figure of merit results after only 180 days of observation, or 800 queries (73\% of the SNPCC spectroscopic sample size). 
Such results were obtained using the same amount of spectroscopic time necessary to observe the original SNPCC spectroscopic sample, showing that this type of strategy is feasible with current available spectroscopic resources. The code used in this work is available in the \href{https://github.com/COINtoolbox/ActSNClass}{COINtoolbox}.
\end{abstract}


\section{Introduction}
\label{sec:intro}

The standard cosmological model rests on three observational pillars: primordial Big-Bang nucleosynthesis \citep{Gamow1948},  the cosmic microwave background radiation \citep{Spergel2007,Planck2016}, and the accelerated cosmic expansion \citep{riess1998,perlmutter1999} - with Type Ia supernovae (SNe Ia) playing an important role in probing the last one. SNe Ia are astronomical transients which are used as standardizable candles
in the determination of extragalactic distances and velocities \citep{hillebrandt2000,goobar2011}. However, between the discovery of a SN candidate and its successful application in cosmological studies, 
additional research steps are necessary.

Once a transient is identified as a potential SN, it must go through three main steps: i) classification, ii) redshift estimation, and iii) estimation of its standardized apparent magnitude at maximum brightness \citep{Phillips93,Tripp98}. Ideally, each SN thus requires at least one spectroscopic observation (preferably around maximum - items i and ii) and a series of consecutive photometric measurements (item iii). Since we are not able to get spectroscopic measurement for all transient candidates, soon after a variable source is detected a decision must be made regarding its spectroscopic follow-up, making coordination between transient imaging surveys and spectroscopic facilities mandatory.  From a traditional perspective, taking a spectrum of a transient that ends up classified as a SNIa results in the object being included in the cosmological analysis. On the other hand, if the target is classified as non-Ia, spectroscopic time for cosmology is essentially considered ``lost''\footnote{This is strictly for cosmological purposes; spectroscopic observations are extremely valuable, irrespective of the transient in question, though for different scientific goals.}.

In the last couple of decades, a strong community effort has been devoted to the detection and follow-up of SNe Ia for cosmology. Classifiers (human or artificial) on which follow-up decisions are based have become increasingly efficient in identifying SNe Ia from early stages of their light curve evolution - successfully targeting them for spectroscopic observations \citep[e.g.][]{Perrett10}. The available cosmology data set has grown from 42 \citep{perlmutter1999} to 740 \citep{betoule2014} in that period of time. This success helped  building consensus around  the paramount importance of SNe Ia for cosmology. It has also encouraged the community to add even more objects to the Hubble diagram and to investigate the systematics uncertainties which currently dominate dark energy error budget \citep[e.g.][]{Conley11}.
Thenceforth,  SNe Ia are major targets of  many current - e.g.  \textit{Dark Energy Survey}\footnote{https://www.darkenergysurvey.org/} (DES),  \textit{Panoramic Survey Telescope and Rapid Response System}\footnote{https://panstarrs.stsci.edu/} (Pan-STARRS) - and upcoming surveys - e.g. \textit{Zwicky Transient Facility}\footnote{http://www.ptf.caltech.edu/ztf/} (ZTF) and \textit{Large Synoptic Survey Telescope}\footnote{https://www.lsst.org/} (LSST). These latter new surveys are expected to completely change the data paradigm for SN cosmology, increasing the number of available light curves by a few orders of magnitude. 

However, to take full advantage of the great potential in such large photometric data sets, we still have to contend with the fact that spectroscopic resources are - and will continue to be - scarce. The majority of photometrically identified candidates will never be followed spectroscopically. Full cosmological exploitation of wide-field imaging surveys necessarily requires a  framework able to infer reliable spectroscopically-derived features (redshift and class) from purely photometric data. Provided a particular transient has an identifiable host, redshift can be obtained before/after the event from the  host observations (spectroscopic or photometric) or even from the light curve itself \citep[e.g.][]{wang2015}.  On the other hand, classification should primarily be inferred from the light curve\footnote{Although, see \citealt{Foley13}.}. This paper concerns itself with the latter. 

Before we dive into the details of SN photometric classification, it is important to keep in mind that, regardless of the method chosen to circumvent these issues,  photometric information will always carry a larger degree of uncertainty than those from the spectroscopic scenario. Photometric redshift estimations are expected to have non-negligible error bars and, at the same time,  any kind of classifier will carry some contamination to the final SNIa sample. Nevertheless, if we manage to keep these effects under control,  we should be able to use photometrically observed SNe Ia to increase the statistical significance of our results. The question whether the final cosmological outcomes surpass those of the spectroscopic-only sample enough to justify  the additional effort is still debatable. Despite a few reports in this direction using real data from the \textit{Sloan Digital Sky Survey}\footnote{\href{http://www.sdss.org/}{http://www.sdss.org/}} \citep[SDSS - ][]{hlozek2012, Campbell13} and Pan-STARRS \citep{jones2017}, the answer keeps changing as different steps of the pipeline are improved and more data become available. Nevertheless, there seems to be a consensus in the astronomical community that we have much to gain from such an exercise.

The vast literature, with suggestions on how to improve/solve different stages of the SN photometric classification pipeline, is a testimony of the positive attitude with which the subject is approached. For more than 15 years the field has been overwhelmed with attempts relying on  a wide range of  methodologies: colour-colour and colour-magnitude diagrams \citep{Poznanski02,Johnson06}, template fitting techniques \citep[e.g.][]{Sullivan06}, Bayesian probabilistic approaches \citep{Poznanski07,Kuznetsova07}, fuzzy set theory \citep{Rodney09}, kernel density methods \citep{newling2011} and more recently, machine learning-based classifiers \citep[e.g.][]{richards2012, ishida2013, karpenka2013, lochner2016, moller2016, charnock2017, dai2017}.

In 2010, the \textit{SuperNova Photometric Classification Challenge} \citep[SNPCC - ][]{kessler2010} summarized the state of the field by inviting different groups to apply their classifiers to the same simulated data set. Participants were asked to classify a set of light curves generated according to the DES photometric sample  characteristics. As a starting point, they were provided with a  spectroscopic sample enclosing $\sim5\%$ of the total data set, and for which class information was disclosed. The organizers posed three main questions: full light curve classification with and without the use of redshift information (supposedly obtained from the host galaxy) and an early epoch classification - where participants were allowed to use only the first 6 observed points from each light curve. The goal of the latter was to access the capability of different algorithms to advise on spectroscopic targeting while the SN was still bright enough to allow it. A total of 10 groups replied to the call, submitting 13 (9) entries for the full light curve classification with (without) the use of redshift information. No submission was received for the early epoch scenario.
%
%

The algorithms competing in the SNPCC were quite diverse, including template fitting, statistical modelling, selection cuts and machine learning-based strategies \citep[see summary of all participants and result in][]{kessler2010}. Classification results were consistent among different methods with no particular algorithm clearly outperforming all the others.  
The main legacy of this initiative however, was the updated public data set made available to the community once the challenge was completed. It became a test bench for experimentation, specially for machine learning approaches \citep{newling2011, richards2012,karpenka2013,ishida2013}.

One particularly challenging characteristic of the SN classification problem, also present in the SNPCC data, is the discrepancy between spectroscopic and photometric samples.  In a supervised machine learning framework, we have no alternative other than to use spectroscopically classified SNe as training. This turns out to be a serious problem, since one of the hypothesis behind most text-book learning algorithms relies on training being representative of target. Due to the stringent observational requirements of spectroscopy, this will never be the case between  spectroscopic and photometric astronomical samples.  But the situation is even more drastic for SNe, where the spectroscopic follow-up strategy  was designed to target as many Ia-like objects as possible.  
Albeit modern low-redshift surveys try to mitigate and counterbalance this effect (e.g. ASASSN\footnote{http://www.astronomy.ohio-state.edu/\~assassin/index.shtml}, iPTF\footnote{https://www.ptf.caltech.edu/iptf}), the medium/high redshift ($z>0.1$) spectroscopic sample is still heavily under-represented by all non-Ia SNe types. Spectroscopic observations are so time demanding, and the rate with which the photometric samples are increasing is so fast, that the situation is not expected to change even with dedicated spectrographs \citep[OzDES - ][]{Childress17}. This issue has  been pointed out by many post-SNPCC machine learning-based analysis \citep[e.g.][]{richards2012,karpenka2013,varughese2015,lochner2016,charnock2017,revsbech2017}. In spite of the general consensus being that one should prioritize  faint objects for spectroscopic targeting,  as an attempt to increase representativeness \citep{richards2012,lochner2016}, the  details on how exactly this should be implemented are yet to be defined.

Thus the question still remains: how do we optimize the distribution of spectroscopic resources with the goal of improving photometric SN identification? Or, in other words, how do we construct a training sample that maximizes accurate classification  with a minimum number of labels, i.e. spectroscopically-classified SNe?
The above question is similar in context to the ones addressed by an  area of  machine learning called  \textit{active learning} \citep{Settles12,Balcan09,Cohn96}. 

Active Learning (AL) iteratively identifies which objects in the target (photometric) sample would most likely improve the classifier if included in the training data - allowing sequential updates of the learning model with a minimum number of labelled instances. 
It has been widely used in a variety of different research fields, e.g. natural language processing \citep{thompson1999}, spam classification \citep{debarr2009}, cancer detection \citep{liu2004} and sentiment analysis \citep{kranjc2015}.
In astronomy, AL has already been successfully applied in multiple tasks: determination of stellar population parameters \citep{solorio2005}, classification of variable stars \citep{richards2012b}, optimization of telescope choice \citep{xia2016}, static supernova photometric classification \citep{dhargupta17}, and photometric redshift estimation \citep{vilalta17}. There are also reports based on similar ideas by \citet{Masters2015,hoyle2016}.

In this work, we show how active learning enables the construction of optimal training datasets for SNe photometric classification, providing observers with a spectroscopic follow-up protocol on a night-by-night basis. 
The framework respects the time evolution of the survey providing a decision process which can be implemented from the first observational nights --avoiding the necessity of adapting legacy data and the consequent translation between different photometric systems.
The methodology herein employed virtually allows any machine learning technique to outperform itself by simply labelling the data (taking the spectrum) in a way that provides maximum information gain to the classifier. As a case study, we focus on the problem of binary classification, i.e. Type Ia vs non-Ia, but the overall structure can be easily generalized for multi-classification tasks.

This paper is organized as follows: section \ref{sec:data} describes the SNPCC data set, emphasizing the differences between spectroscopic and photometric samples.  Section \ref{sec:preproc} details on the feature extraction, classifier and metrics used throughout the paper.  Section \ref{sec:AL} explains the AL algorithm, details the configurations we chose to approach the SN photometric classification problem, and presents results for the idealized static, full light curve scenario. Section \ref{sec:TD} explains how our approach deals with the real-time evolution of the survey and its consequent results. Section \ref{sec:semi} presents our proposal for real-time multiple same-night spectroscopic targeting (batch-mode AL). Having established a baseline data driven spectroscopic strategy, in Section \ref{sec:tel} we estimate the amount of telescope time necessary to observe the AL queried sample. Lastly, section \ref{sec:conclusions} shows our conclusions.

\section{Data}
\label{sec:data}

\begin{figure}
\centering
\includegraphics[width=\columnwidth, trim={0.5cm 0.5cm 2.5cm 0.0cm}, clip]{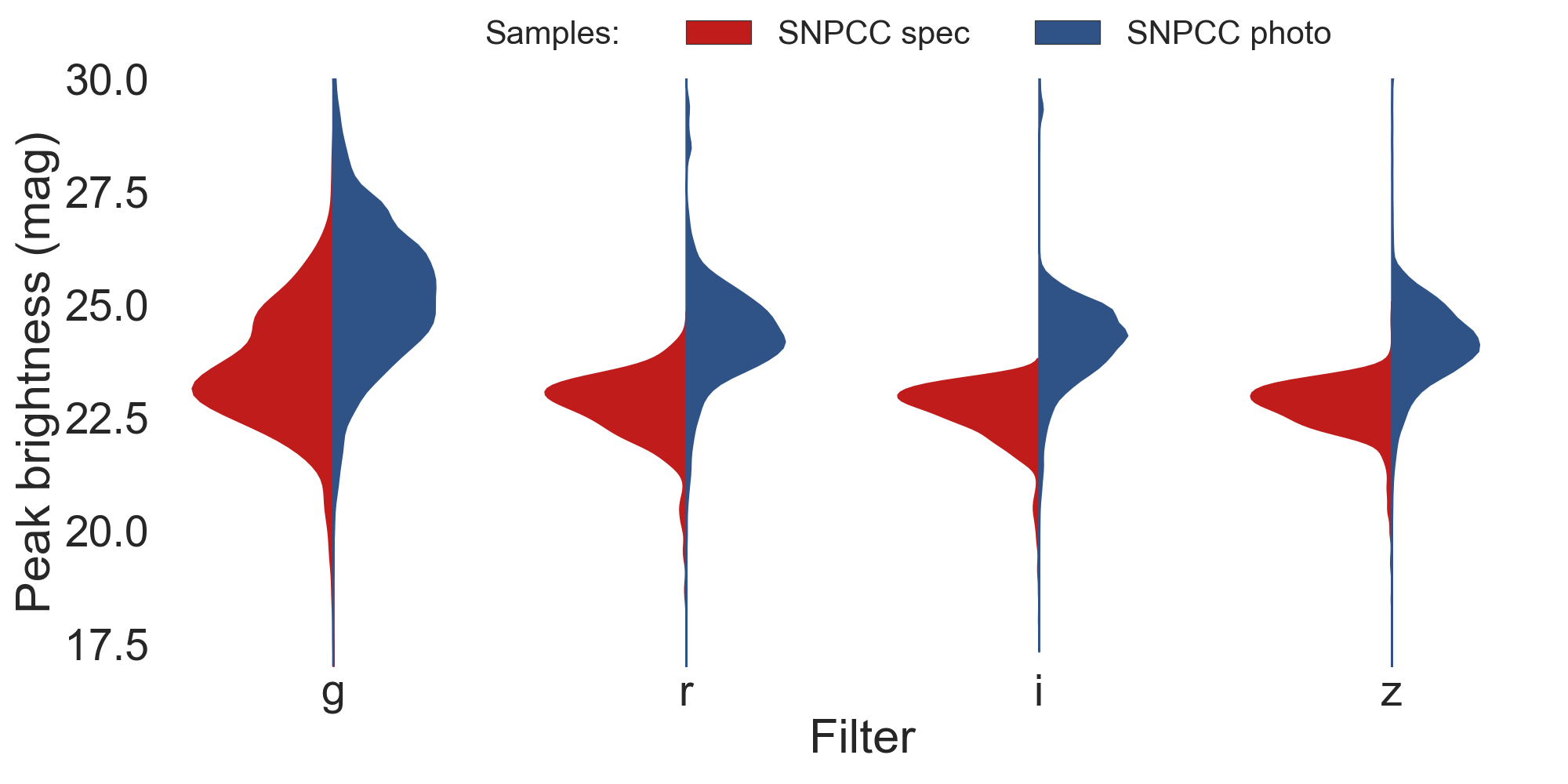}
\caption{Comparison between simulated peak magnitudes in the SNPCC spectroscopic  (red - training) and photometric (blue - target) samples. Violin plots show both distributions in each of the DES filters.}
\label{fig:peakmag}
\end{figure}

\begin{figure}
\centering
\includegraphics[width=\columnwidth]{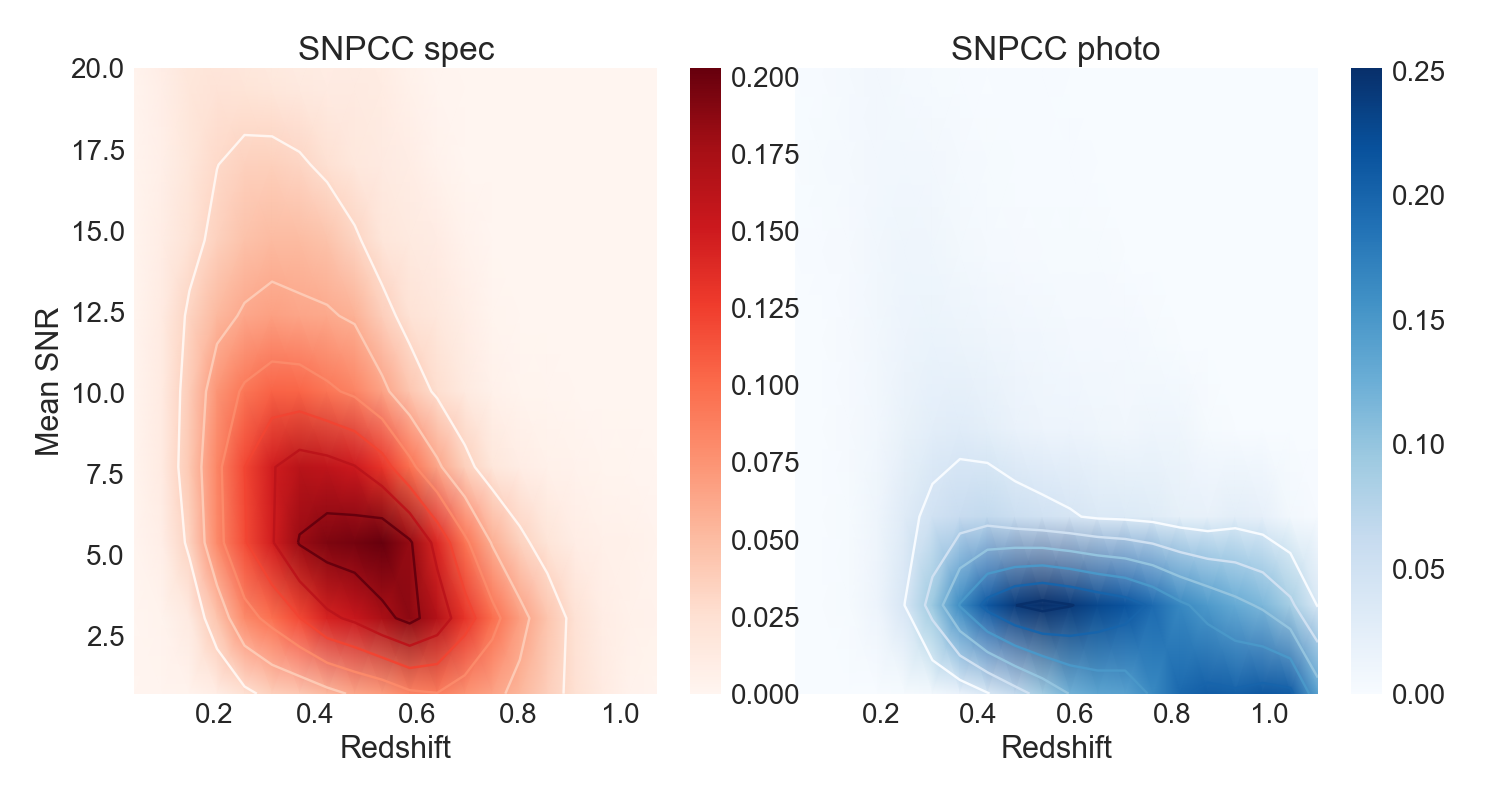}
\caption{Distribution of mean signal to noise ratio (SNR) in the SNPCC spectroscopic (red - training) and photometric (blue - target) samples.}
\label{fig:SNR}
\end{figure}

\begin{figure}
\centering
\includegraphics[width=\columnwidth]{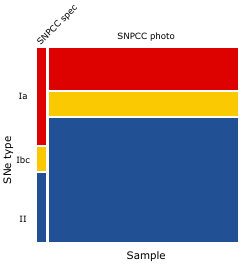}
\caption{Populations of different supernova types in the SNPCC spectroscopic (training) and photometric (target) samples. The spectroscopic sample holds 599 (51\%) Ia, 144 (13\%) Ibc and 400 (36\%) II, while the photometric sample comprises 4326 (22\%) Ia, 2535 (13\%) Ibc and 12442 (65\%) II.}
\label{fig:types_ini}
\end{figure}

In what follows, we use the data released after the SNPCC. This is a simulated data set constructed to mimic DES observations.
The sample contains 20216 supernovae (SNe) observed in four DES filters, $\{g,r,i,z\}$, among which a subset of 1103 are identified as belonging to the \textit{spectroscopic} sample. This subset was constructed considering observations through a 4m (8m)  telescope and limiting $r$-band magnitude of 21 (23.5) \citep{kessler2010}. Thus, it resembles closely  biases foreseen in a realistic spectroscopic sample when compared to the photometric one. Among them, we highlight the predomination of brighter objects (figure \ref{fig:peakmag}) with higher signal to noise ratio (SNR, figure \ref{fig:SNR}), and the predominance  of SNe Ia over other SN types (figure \ref{fig:types_ini}). 
Hereafter, the spectroscopic sample will be designated \textit{SNPCC spec} and the remaining objects will be addressed as \textit{SNPCC photo}.

\section{Traditional analysis}
\label{sec:preproc}

\subsection{Feature extraction}
\label{subsec:fit}

For each supernova, we observe its light curve, i.e. the evolution of  brightness (flux) as a function of time, in four DES filters $\{g,r,i,z\}$. 
For most machine learning applications, this information needs to be homogenized before it can be used as input to a learning algorithm\footnote{Exceptions include algorithms able to deal with a high degree of missing data \citep[e.g.][]{charnock2017, Naul2018}.}. There are many ways in which this feature extraction step can be performed: via a proposed analytical parametrization  \citep{Bazin09,newling2011}, comparisons with theoretical and/or well-observed templates \citep{sako2008} or dimensionality reduction techniques \citep{richards2012,ishida2013}. Literature has many examples showing that, for the same classification algorithm, the choice of the feature extraction method can significantly impact classification results \citep[see][and references therein]{lochner2016}. 

\begin{figure}
\centering
\includegraphics[width=\columnwidth, trim={1.0cm 1.0cm 1.0cm 0.5cm}]{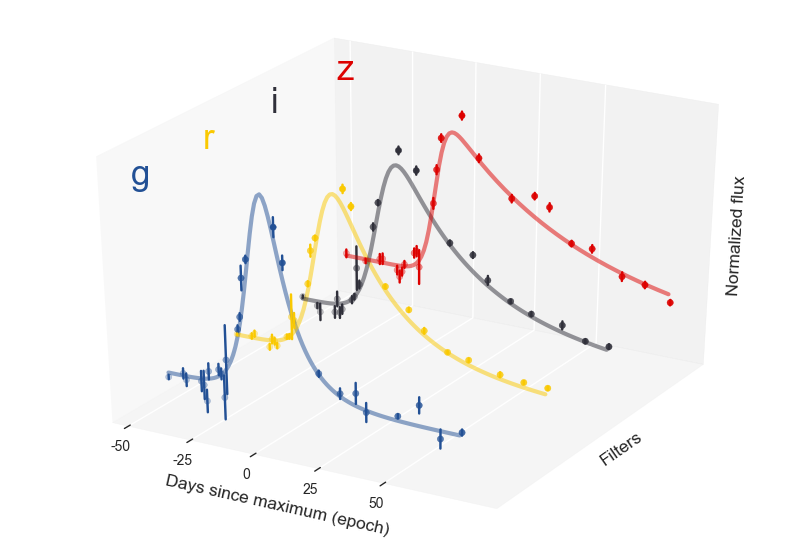}
\caption{Example of light curve fitted with the parametric function of \citet{Bazin09} - equation \ref{eq:bazin}. The plot shows measurements for a typical well-sampled type Ia at redshift $z\sim0.20$, in each one of the 4 DES filters (dots and error bars) as well as its best fitted results (lines).}
\label{fig:lc_fit}
\end{figure}

In what follows, we use the parametrization proposed by \citet{Bazin09},

\begin{equation}
f(t) = A\frac{e^{-(t-t_0)/\tau_f}}{1+e^{(t-t_0)/\tau_r}}+B,
\label{eq:bazin}
\end{equation}
\noindent
where $A$, $B$, $t_0$, $\tau_f$ and $\tau_r$ are parameters to be determined.
We fit each filter independently in flux space with a Levenberg-Marquardt least-square minimization \citep{LevenbergMarquardt}.  
Figure \ref{fig:lc_fit} shows an example of flux measurements, corresponding errors and best-fit results in all 4 filters for a typical, well-sampled, SN Ia from SNPCC data.

Although not optimal for such a diverse light curve sample, the parametrization given by Eq. \ref{eq:bazin} was chosen for being a fast feature extraction method. Moreover, as any parametric function, it returns the same number of parameters independently of the number of observed epochs, which is crucial for dealing with an inhomogeneous time-series which changes on a daily basis (the importance of this homogeneous representation is further detailed in Section \ref{sec:TD}). We stress that a more flexible feature extraction procedure still holding the characteristics described above would only improve the results presented here.

\subsection{Classifier}
\label{subsec:RF}

Once the data has been homogenized, we need a supervised learning model to harvest the information stored in the spectroscopic sample. Analogous to the feature extraction case, the choice of classifier also impacts the final classification results for a given static data set \citep{lochner2016}. 
In order to isolate the impact of AL in improving a given configuration of feature extraction and machine learning pipeline, we chose to restrict our analysis to a single classifier. A complete study on how different classifiers respond to the update in training provided by AL is out of the scope of this work, but is a crucial question to be answered in subsequent studies. All the results we present below were obtained with a \textit{random forest} algorithm \citep{Breiman01}. 

Random forest is a popular machine learning algorithm known to achieve accurate results with minimal parameter tuning. It is an ensemble technique made up of multiple decision trees \citep{breiman1984}, constructed over different sub-samples of the original data. Final results are obtained by averaging over all trees \citep[for further details, see appendices A and B of][]{richards2012}. The method has been successfully used for SN photometric classification \citep{richards2012,lochner2016,revsbech2017}. 
In what follows, we used the \textsc{scikit-learn}\footnote{\href{http://scikit-learn.org/}{http://scikit-learn.org/}} implementation of the algorithm with 1000 trees. In this context, the probability of being a SN Ia, $p_{\mathrm{Ia}}$, is given by the percentage of trees in the ensemble voting for a SNIa classification\footnote{In this work we are concerned only with Ia $\times$ non-Ia classification. The analysis of classification performance using other SN types will be the subject of a subsequent investigation.}. 

\subsection{Metrics}
\label{subsec:metrics}

The choice of a metric to quantify classification success goes beyond the use of classical accuracy (equation \ref{eq:acc}) - especially when the populations are unbalanced (figure \ref{fig:types_ini}). In order to optimize information extraction, this choice must take into account the scientific question at hand. 

In the traditional SN case, the goal is to improve the quality of the final SNIa sample for further cosmological use. In this context, a false negative (a SNIa wrongly classified as non-Ia) will be excluded from further analysis posing no damage on subsequent scientific results. On the other hand, a false positive (non-Ia wrongly classified as a Ia) will be mistaken by a standard candle, biasing the cosmological analysis. Thus, purity (equation \ref{eq:pur}) of the photometrically classified SNIa set is of paramount importance. At the same time, we wish to identify as many SNe Ia as possible (high efficiency - equation \ref{eq:eff}), in order to guarantee optimal exploitation of our resources. Taking such constraints into consideration, \citet{kessler2010} proposed the use of a figure of merit which penalizes classifiers for false positives (equation \ref{eq:fom}). 
Throughout our analysis, classification results will be reported according to these 4 metrics: 

\begin{equation}
\textrm{accuracy} = \frac{N_{\textrm{sc}}}{N_{\textrm{tot}}}, 
\label{eq:acc}
\end{equation}
\begin{equation}
\textrm{efficiency} =  \frac{N_{\textrm{sc,Ia}}}{N_{\textrm{tot,Ia}}}, \label{eq:eff}
\end{equation}
\begin{equation}
\textrm{purity} =  \frac{N_{\textrm{sc,Ia}}}{N_{\textrm{sc,Ia}} + N_{\textrm{wc,nIa}}}, \qquad \textrm{and}  \label{eq:pur}
\end{equation}
\begin{equation}
\textrm{figure of merit} = \frac{N_{\textrm{sc,Ia}}}{N_{\textrm{tot,Ia}}} \times \frac{N_{\textrm{sc,Ia}}}{N_{\textrm{sc,Ia}} + W N_{\textrm{sc,nIa}}}\label{eq:fom}, 
\end{equation}

\noindent where $N_{\textrm{sc}}$ is the toal number of successful classifications, $N_{\textrm{tot}}$ the total number of objects in the target sample, $N_{\textrm{sc,Ia}}$  the number of successfully  classified SNe Ia (true positives), $N_{\textrm{tot,Ia}}$ the total number of SNe Ia in the target sample, $N_{\textrm{wc,nIa}}$ the number of non-Ia SNe wrongly classified as SNe Ia (false positives) and $W$ is a factor which penalizes the occurrence of false positives. Following \citet{kessler2010} we always use $W=3$.

In the AL framework we propose, the metrics above are used to quantify the classification results in the target sample. They were calculated after the classifications were performed and had no part in the decision making algorithm (further details in Section \ref{sec:AL} and Appendix \ref{ap:AL}).

\section{Active Learning}
\label{sec:AL}

We now turn to the key missing ingredient in our pipeline. The tools we have described thus far allow us to process (section \ref{subsec:fit}), classify (section \ref{subsec:RF}), and evaluate classification results (section \ref{subsec:metrics}) given a pair of labelled and unlabelled light-curve data sets. The question now is: starting from this initial configuration, how can we optimize the use of subsequent spectroscopic resources in order to maximize the potential of our classifier? Or in other words, how can we achieve high generalization performance by adding a minimum number of new spectroscopically observed objects to the training sample? We advocate the use of a dedicated recommendation system tuned to choose the most informative objects in a given sample - those which will be spectroscopically targeted. 

\begin{figure}
\centering
\includegraphics[width=\columnwidth]{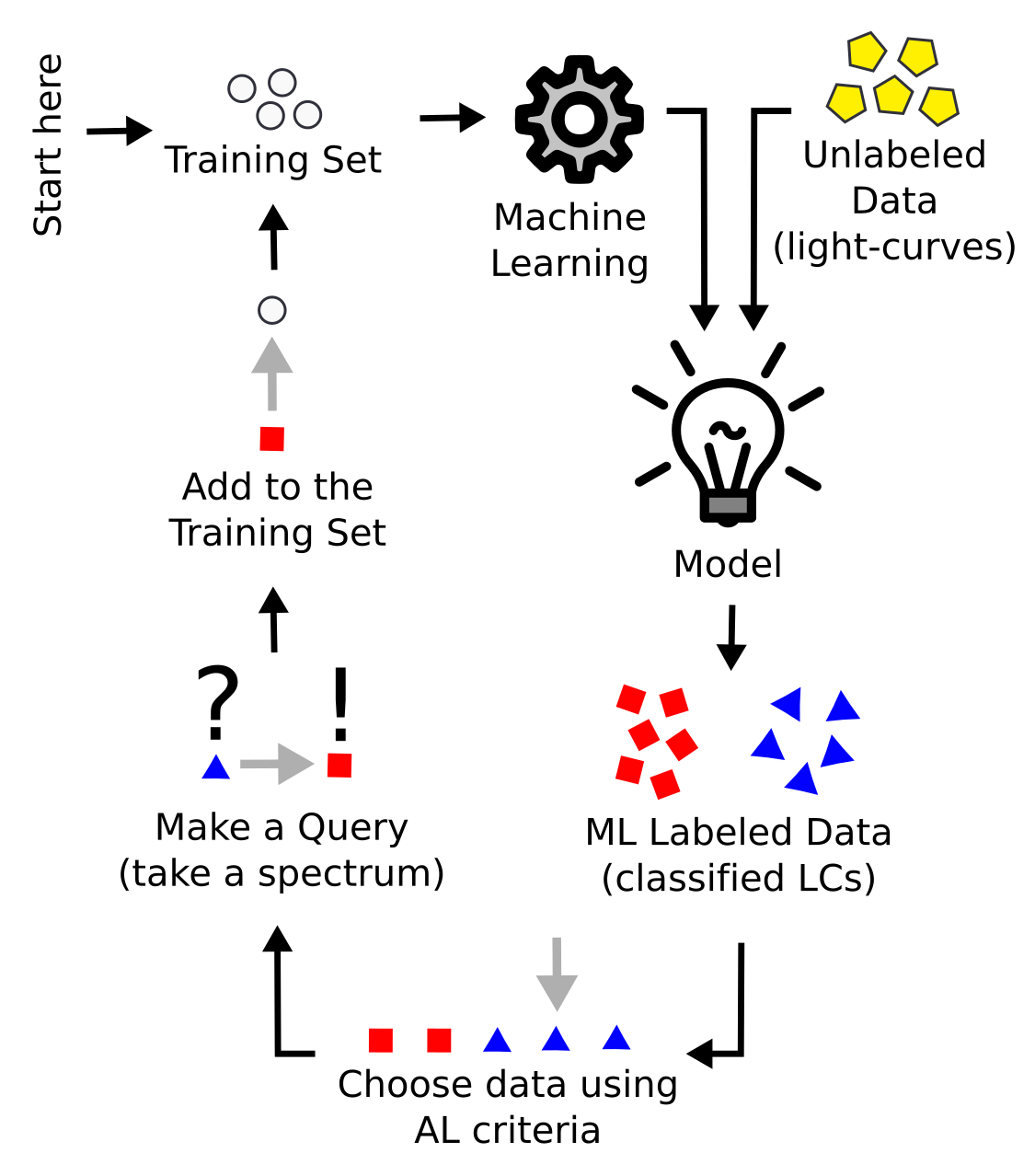}
\caption{Schematic illustration of the Active Learning (AL) work flow in the context of photometric light curves classification. Starting at the top left, the training set (spectroscopic sample - grey circles), is used to train a  machine learning algorithm - resulting in  a model which is then applied to the unlabelled data (photometric light-curves - yellow pentagons). This initial model returns a classification for each data point of the unlabelled set (now represented as red squares and blue triangles). The AL algorithm is then used to choose a data point of the unlabelled data with highest potential to improve the classification model (identified by the grey arrow). The label of this point is then queried (a spectrum is taken).  Once the true label of the queried point is known, it is added to the training set (converted into a grey circle), and the process is repeated.} 
\vspace*{-5mm}
\label{fig:AL_diagram}
\end{figure}

Active learning (AL) is an area of machine learning designed to optimize learning results while minimizing the number of required labelled instances. At each iteration, the algorithm suggests which of the unlabelled objects would be most informative to the learning model if its label was available \citep{Settles12,Balcan09,Cohn96}.  Once identified, a query is made - in other words, the algorithm is allowed to interact with an external agent in order to ask for the label of that object\footnote{In the machine learning jargon, this agent is called an \textit{oracle}. It can be a human, machine or piece of software capable of providing labels. In our context, making a query to the oracle corresponds to spectroscopically determining the class of a given object.}. The queried object - along with its label - is then 
added to the training sample and the model is re-trained. This process is repeated until convergence is achieved, or until labelling resources are exhausted. The complete algorithm is illustrated in Fig. \ref{fig:AL_diagram}.

Different flavours of AL propose different strategies to identify which objects should be queried. In what follows, we report results obtained using pool-based AL\footnote{Our analysis used the \textsc{libact} Python package developed by \citet{yang2017}.} via uncertainty sampling.  In this framework, we start with two sets of data: labelled and unlabelled. At first, the  machine learning algorithm (classifier) is trained using all the available labelled data. Then, it    is used to provide a class probability for all objects in the unlabelled set. The object whose classification is most uncertain is chosen to be queried. In a binary classification problem as the one investigated here, this corresponds to querying objects near the boundary between the two classes - where the classifier is less reliable. 
A detailed explanation of the uncertainty sampling technique (as well as \textit{query by committee}) is given in appendix \ref{ap:AL}.

Finally, whenever one wishes to quantify the improvement in classification results due to AL, it is important to keep in mind that simply increasing the number of objects available for training changes the state of the model - independently of how the extra data were chosen. Thus, AL  results should always be compared to the \textit{passive learning} strategy, where at each iteration objects to be queried are randomly drawn from the unlabelled sample. Moreover, in the specific case of SN classification, we also want to investigate how the learning model would behave if the same number of objects were queried following the \textit{canonical} spectroscopic targeting strategy  - where objects to be queried are randomly drawn from a sample which follows closely the initial SNPCC labelled distribution. Diagnostic results presented bellow  show outcomes from all three strategies: canonical, passive learning, and AL via uncertainty sampling\footnote{The code used in this paper can be found in the COINtoolbox - \href{https://github.com/COINtoolbox/ActSNClass}{https://github.com/COINtoolbox/ActSNClass}}.

\subsection{Static full light curve analysis}
\label{subsec:fullLC}

We begin by applying the complete framework described in the previous subsections to static data. This is the traditional approach, where we consider that all light curves were completely observed at the start of the analysis. Although this is not a realistic scenario (one cannot query, or spectroscopically observe,  a SN that has already faded away), it gives us an upper limit on estimated classification results. Section \ref{sec:TD} considers more realistic constraints on available query data and light-curve evolution.

\begin{figure*}
\centering\includegraphics[width=\textwidth]{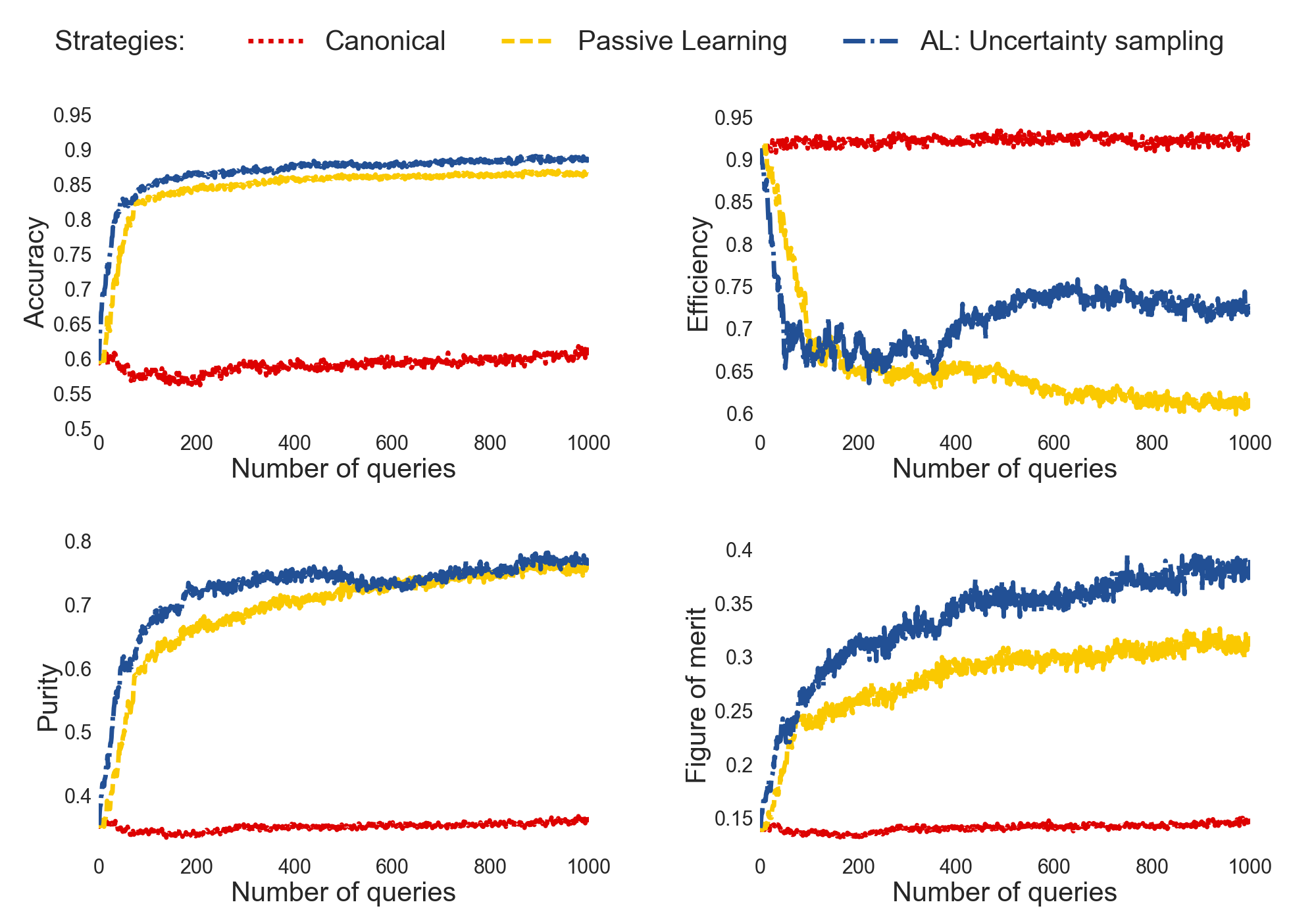}
\caption{Evolution of classification results as a function of the number of queries for the static full light curve analysis.}
\label{fig:fullLC}
\end{figure*}

For each light curve and filter, all available data points were used to find the best-fit parameters of equation \ref{eq:bazin} following the procedure described in section \ref{sec:preproc}. Best-fit values for different filters were concatenated to compose a line in the data matrix. In order to ensure the quality of fit, we considered only SNe with a minimum of 5 observed epochs in each filter; this reduced the size of our spectroscopic and photometric samples to 1094 and 20193 objects, respectively. 

\subsubsection{Sub-samples}
\label{subsubsec:subsamples}

The iterative framework presented above corresponds to the AL strategy for choosing the next object to be queried. In this description, we have 2 samples: labelled and unlabelled. 
In case we wish to quantify the performance of the ML algorithm after each iteration, the recently re-trained model must be used to predict the classes of objects in a third sample --one that did not take part in the AL algorithm. Classification metrics are then calculated, after each iteration, from predictions on this independent sample. In this scenario we need 3 samples: training, query, and target. The \textit{query sample} corresponds to  the set of all objects \textit{available for query} upon which the model evolves\footnote{Not to be confused with the \textit{set of queried objects}, which comprises the specific objects added to the training set (1 per iteration).}. On the other hand, the \textit{target sample} corresponds to the independent one over which diagnostics are computed. In the traditional analysis, query and target sample follow the same underlying distribution in feature space; this separation helps avoid overfitting. This is the case for the static full light curve analysis. In the results presented in this sub-section, \textit{SNPCC photo}  was randomly divided into query (80\%) and target (20\%) samples. 

Finally, we quantified the evolution of the classification results when new objects are added to the training sample according to the canonical spectroscopic follow-up strategy, by constructing a \textit{pseudo-training} sample. For each element of \textit{SNPCC spec}, we searched for its nearest neighbour in \textit{SNPCC photo}\footnote{This calculation was performed in a 16 dimensions parameter space: type, redshift, simulated peak magnitude, and mean SNR in all 4 filters. For all the numerical features we used a standard euclidean distance.}. This allowed us to construct a data set which follows very closely the distribution in the parameter space covered by the original \textit{SNPCC spec}. Thus, randomly drawing elements to be queried from this \textit{pseudo-training} sample is equivalent to feeding more data to the model according to the canonical spectroscopic follow-up strategy.

\begin{figure*}
\centering
\includegraphics[width=\textwidth]{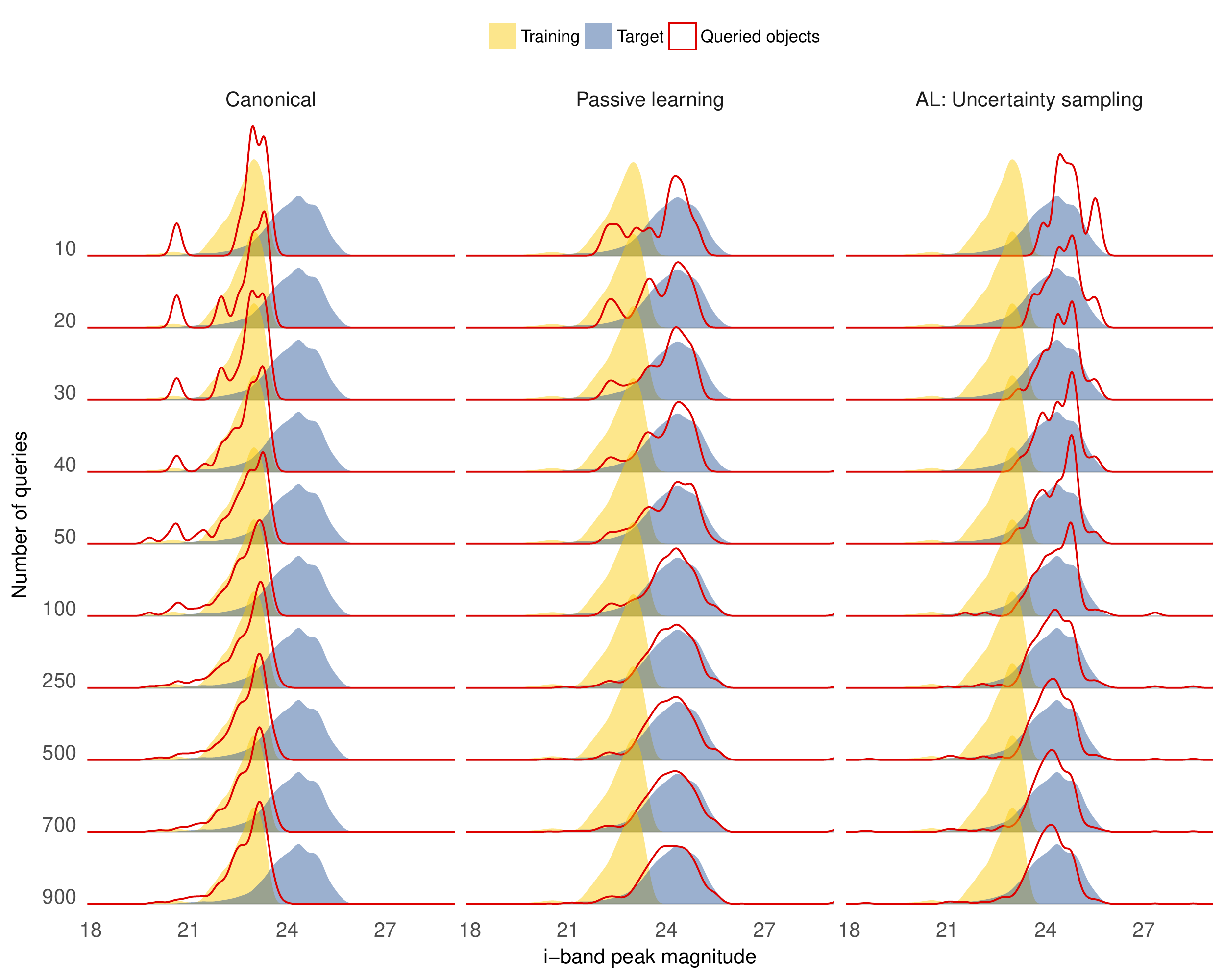}
\caption{Simulated $i$-band peak magnitude distribution as a function of the number of queries for the static full light curve scenario. The yellow (blue) region shows distribution for the training (target) samples, while the red curves denote the composition of sample queried by AL. Lines go through 10 to 900 queries (from top to bottom). Different columns correspond to different learning strategies: canonical, passive and active learning via uncertainty sampling (from left to right).}
\label{fig:joy_fullLC}
\end{figure*}

\begin{figure*}
\centering
\includegraphics[width=\textwidth]{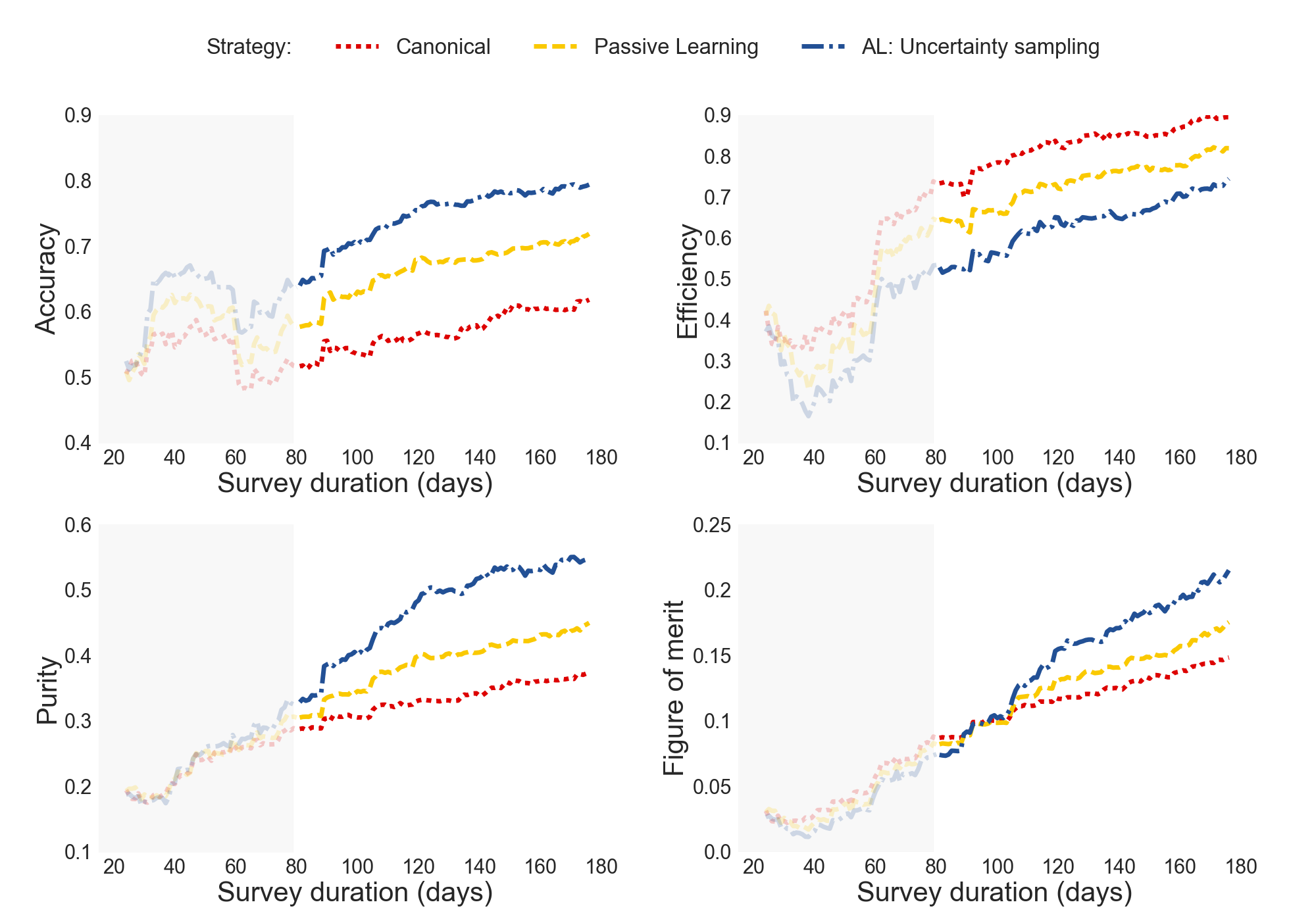}
\caption{Evolution of the classification results as a function of the survey duration for the time-domain AL considering the SNPCC training set as completely given in the beginning of the survey.}
\label{fig:TD1}
\end{figure*}

\subsubsection{Results}
\label{subsubsec:res_fullLC}

In this section we present classification results for the static full light curve scenario according to 3 spectroscopic targeting strategies: canonical, passive learning  and AL via uncertainty sampling (section \ref{sec:AL}). In all three cases, at each iteration 1 object was queried and added to the training sample (the one with highest uncertainty). We allowed a total of 1000 queries, almost doubling the original training set. 

Figure~\ref{fig:fullLC} shows how classification diagnostics evolve with the number of queries.  The red inverse triangles describe results following the canonical strategy (random sampling from the pseudo-training sample), yellow circles show results from passive learning (random sampling from the query sample), and blue triangles represent results for AL via uncertainty sampling. We notice that  the canonical spectroscopic targeting strategy does not add new information to the model --even if more labelled data is made available. Thus there is almost no change in diagnostic results after 1000 queries. On the other hand, the canonical strategy is very successful in identifying SN Ia (approximately 92\% efficiency); however, by prioritizing  bright events, it fails to provide the model with enough information about other SN types. Consequently, its performance in other diagnostics is poor ($\sim 60\%$ accuracy, 36\% purity and a figure of merit of 0.15). At the same time, passive learning and AL via uncertainty sampling show very similar efficiency results up to 400 queries. Accuracy levels stabilize quickly (84\%/87\% after only 200 queries), followed closely by purity results (73\% after 600 queries). The biggest difference appears on efficiency levels. We can recognize an initial drop in efficiency up to 400 queries. This is expected, since both strategies prioritize the inclusion of non-Ia objects in the training sample: passive learning simply led by the higher percentages of non-Ia SNe in the target sample (figure \ref{fig:types_ini}), and AL by aiming at a more diverse information pool.  This implies that high accuracy and purity levels are accompanied by a decrease in efficiency (from 92\% to 68\% at 200 queries).  After a minimally diverse sample is gathered, passive learning continues to lose efficiency, stabilizing at 63\% after 700 queries, while AL is able to harvest further information to stabilize at 72\% after 800 queries. Thus, after 1000 new objects were added to the training sample, passive learning achieves a figure of merit of 0.32 (2.1 times higher than canonical), while AL via uncertainty sampling achieves a figure of merit of 0.39  (2.6 times higher than canonical).

Figure \ref{fig:joy_fullLC} illustrates how the distribution of peak $i$-band magnitude in the set of queried elements evolves with the number of queries. 
For the sake of comparison, we also show the static distributions for the training (yellow) and target (blue) samples. As expected, the canonical strategy consistently follows the spectroscopic sample distribution. Meanwhile, passive learning  completely ignores the existence of the initial training --consequently, its initial queries overlap with regions already covered by the training sample, allocating a significant fraction of spectroscopic resources to obtain information already available in the training. The AL strategy, even in very early stages, takes into account the existence of the training sample, focusing its queries in the region not covered by training data (higher magnitudes). At 900 queries, the set of queried objects chosen by passive learning (red line, middle column) follows closely the distribution found in the  target sample (blue), - but this does not translate into a better classification because the bias present in the original training was not yet overcome. On the other hand, the discrepancy in distributions between the target sample (blue region) and the set of objects queried by AL (red line, right-most column) at 900 queries is a consequence of the existence of the initial training\footnote{The reader should keep in mind that after 1000 queries the model is trained in  a sample containing the complete SNPCC spectroscopic sample added to the set of queried objects.}. The fact that AL takes this into account is reflected in the classification results (figure \ref{fig:fullLC}). 

These results provide evidence that AL algorithms are able to improve SN photometric classification results over canonical spectroscopic follow-up strategies, or even passive learning in a highly idealized environment\footnote{A result already pointed out by \citet{dhargupta17}.}. However, in order to have a more realistic description of a SN survey, we need to take into account the transient nature of the SNe and the evolving aspect of an observational survey. 

Although we chose to illustrate non-representativeness between samples in terms of peak brightness in different bands (e.g. figures \ref{fig:peakmag}, \ref{fig:joy_fullLC} and \ref{fig:peakmag_samples}), these features are absence in the input data matrix. Our goal is to emphasize that the underlying astrophysical properties are tracked differently by the AL and passive learning strategies - even if these are not explicitly used.

\section{Real-time analysis}
\label{sec:TD}

In this section, we present an approach to deal with the time evolving aspect of spectroscopic follow-ups in SN surveys. This is done through the daily update of:
\begin{enumerate}
\item identification of objects allocated to query and target samples,
\item feature extraction and
\item model training.
\end{enumerate}

\begin{figure}
\centering
\includegraphics[width=\columnwidth]{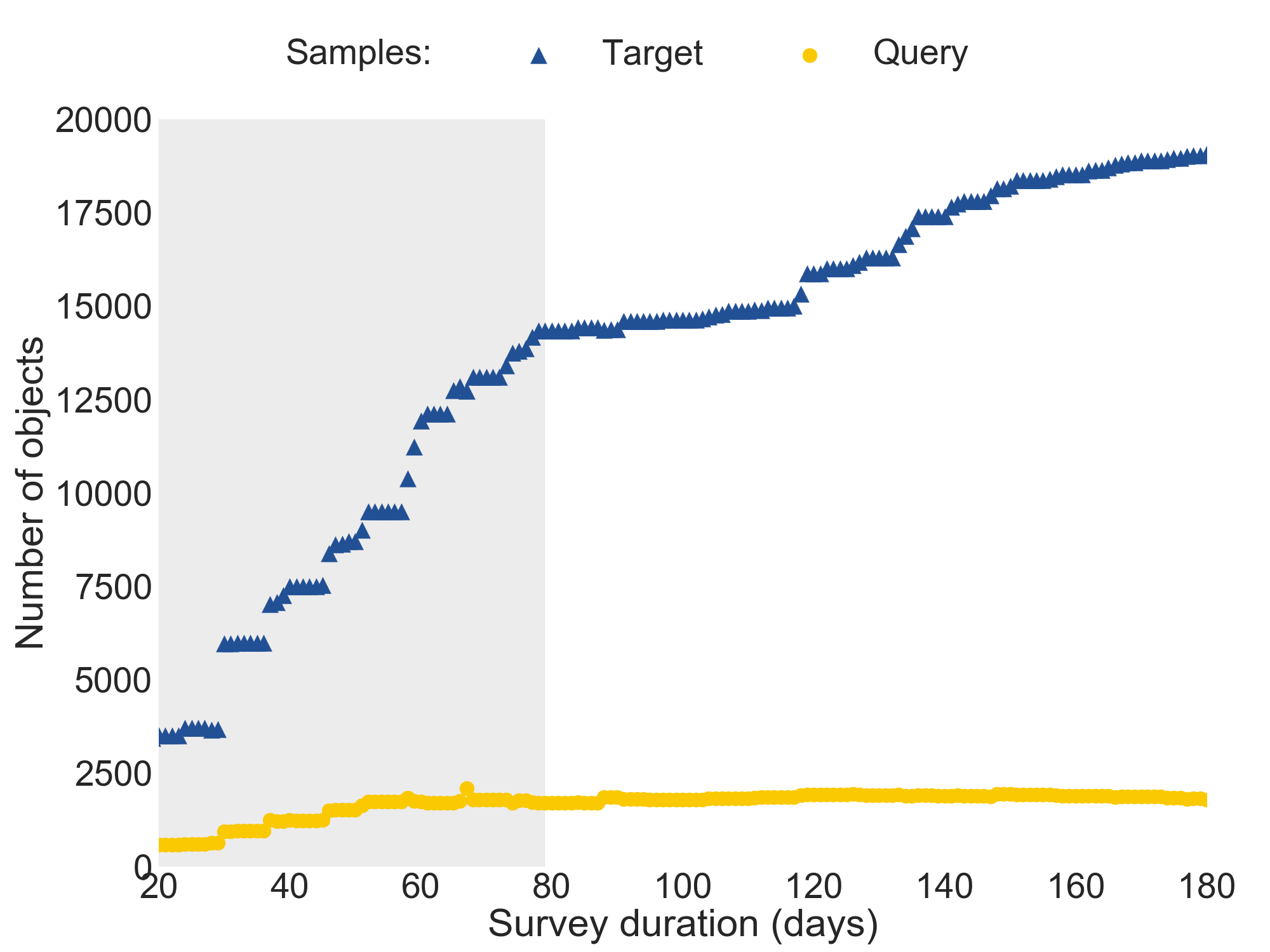}
\caption{Number of objects in the query (yellow circles) and target (blue triangles) samples as a function of the days of survey duration. The grey region highlights the initial build-up phase of the survey, where there is a steep increase in the number of objects in the target sample.}
\label{fig:samples}
\end{figure}

We begin considering the full SNPCC spectroscopic sample completely observed at the beginning of the survey - this allows us to have an initial learning model. 
Then, at each observation day $d$, a given SN is included in the analysis if, until that moment, it has at least 5 observed epochs in each filter. If this first criterion is fulfilled,  the object is designated as part of the \textit{query sample} if its  $r$-band magnitude is lower than or equal to 24 ($m_{r}\leq 24$ at $d$) - otherwise, it is assigned to the target sample\footnote{We consider an object with $r$-band magnitude of 24 to have the minimum brightness necessary to allow spectroscopic observation with a 8-meter class telescope.}. Figure \ref{fig:samples} shows how the number of objects in the query (yellow circles) and target (blue triangles) samples evolves as a function of the number of observing days. Although the survey starts observing at day 1, we need to wait until day 20 in order to have at least 1 object with a minimum of 5 observed epochs in each filter. From then on, the query sample  begins with 666 objects (at day 20) and shows a steady increase until it almost stabilizes $\sim 2100$ SNe (around day 60).  On the other hand, the target sample shows a steep increase until $d \sim 80$ (hereafter, \textit{build-up phase}) and continuous to grow from there until the end of the survey - although at  a lower rate. This behaviour is expected since, in this description,  the \textit{query sample} corresponds to the fraction of photometric objects satisfying the magnitude threshold ($m_{r} \leq 24$)  at a specific time.  Notice that as the survey evolves, an object whose detection happened in a very  early phase will be assigned to the target sample during its rising  period, but if its brightness increases enough to allow spectroscopic targeting it will move to the  query sample - where it will remain for a few epochs. After its maximum passes, the SN will eventually return to the target sample as soon as it fades below the magnitude threshold - remaining there until the end of the survey. Thus, it is important to keep in mind that, despite its size being practically constant after the build-up phase, individual objects composing the query sample might not be the same for consecutive days. 

The feature extraction process is also performed on a daily basis, considering only  the epochs measured until that day. This clarifies why we consider an analytical parametrization a simple, and efficient enough, feature extraction procedure.  It reasonably fast and  encompasses prior domain knowledge on light curve behaviour while returning the same number of parameters independently of the number of observed epochs. Moreover, it avoids the necessity of determining the time of maximum brightness or performing any type of epoch alignment \citep[see e.g.][]{richards2012,ishida2013,revsbech2017}. Thus, we are able to update the feature extraction step as soon as a new epoch is observed and still construct a homogeneous and complete low-dimensionality data matrix. The only constraint is the number of observed epochs, which  must be at least equal to the number of parameters in all filters. 

\begin{figure*}
\centering
\includegraphics[width=\textwidth]{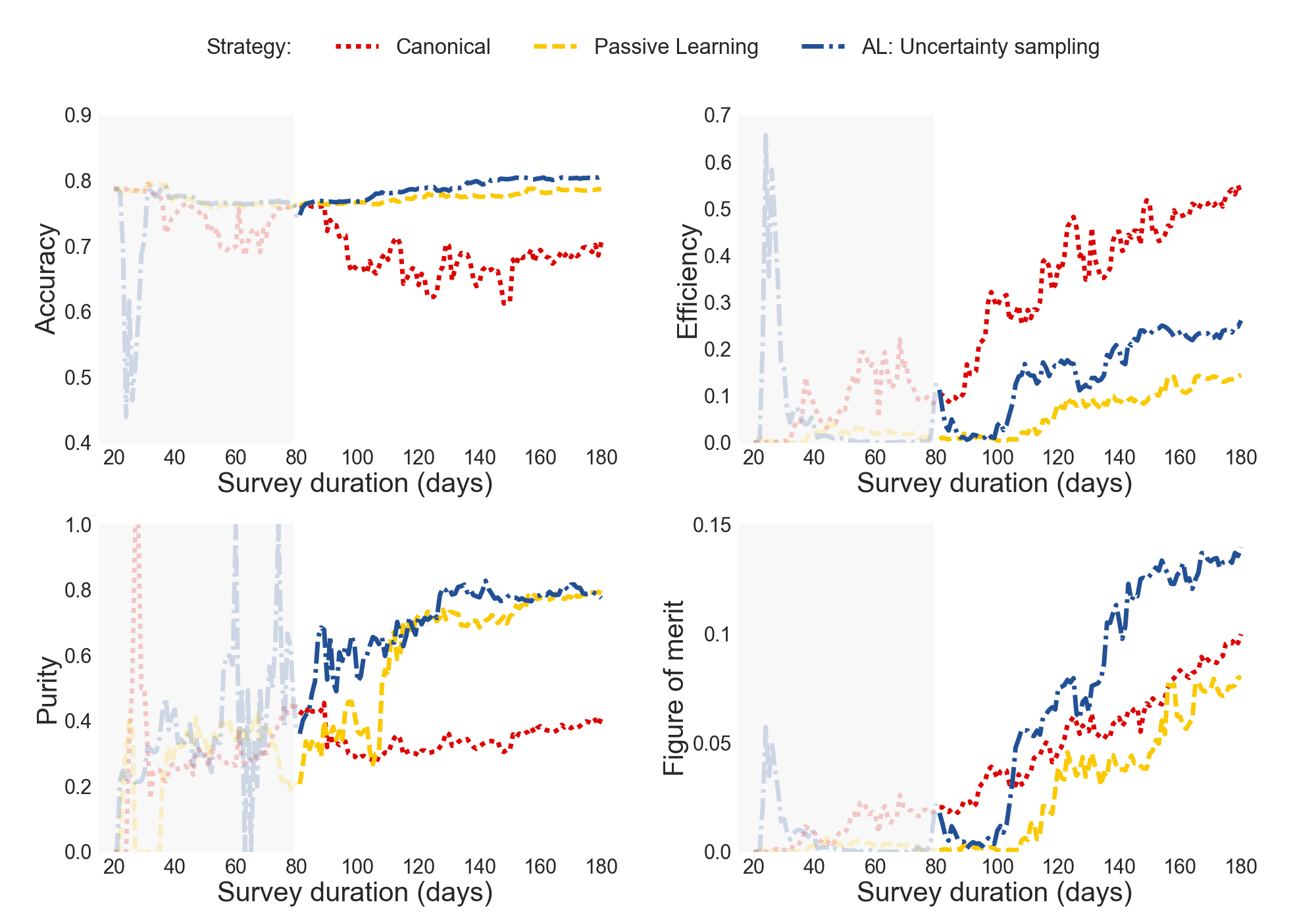}
\caption{Evolution of classification results as a function of survey duration in the real time analysis, with a random initial training of 1 object.}
\label{fig:TD2}
\end{figure*}

Finally, at the end of each night, the model is trained using the features and labels available until that point. The AL algorithm is allowed to query only the objects belonging to the \textit{query sample}. Once a query is made, the targeted object and corresponding label are added to the training sample, the model is re-trained and the result applied to the target sample (figure  \ref{fig:AL_diagram}).  Given the time span of the SNPCC data, we are able to repeat this analysis  for a total of  180 days.

\begin{figure*}
\centering
\includegraphics[width=\textwidth]{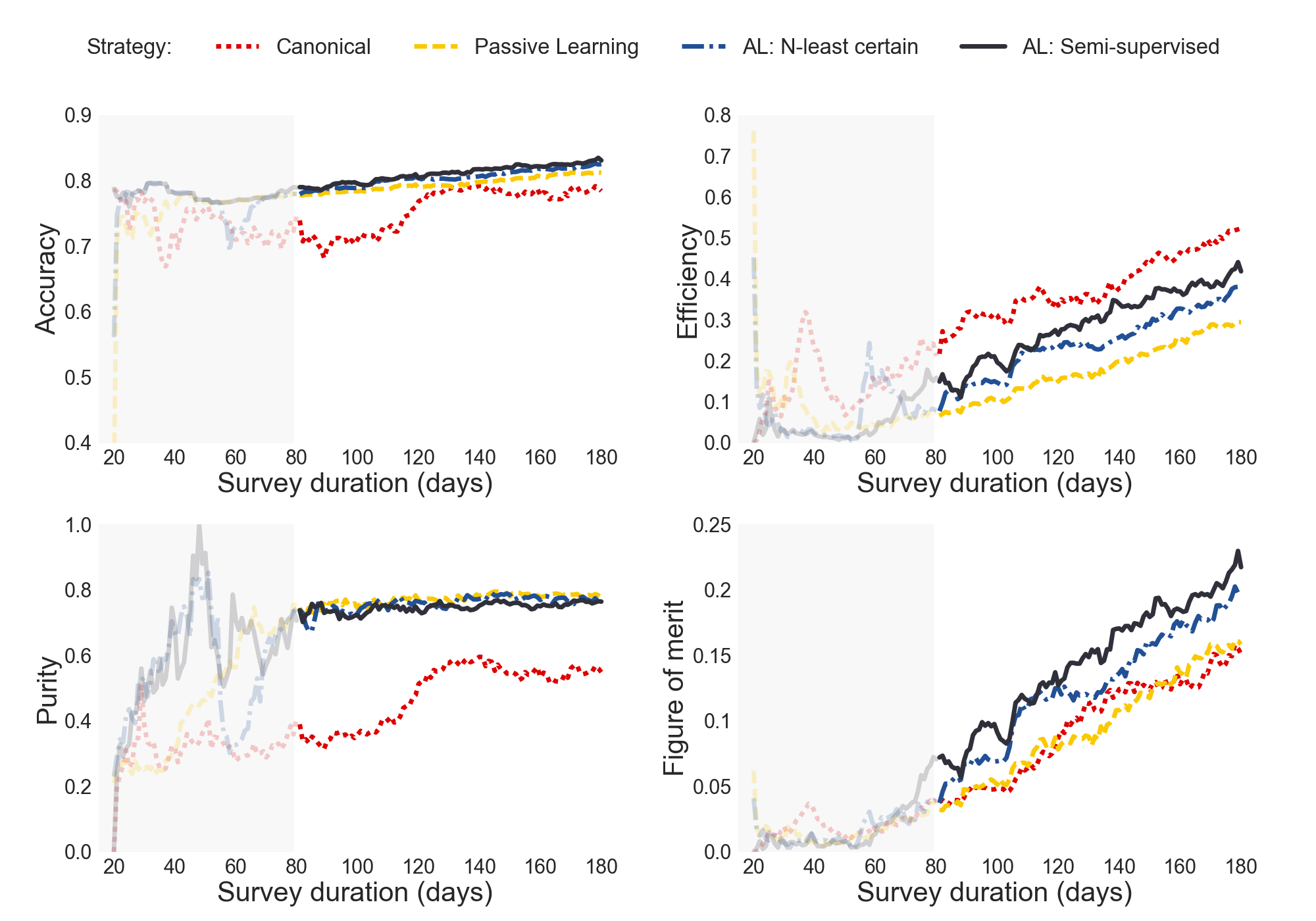}
\caption{Evolution of classification results as a function of survey duration for the batch-mode real time analysis with $N=5$ and a random initial training of 5 objects.}
\label{fig:TD_batch5}
\end{figure*}

Figure \ref{fig:TD1} shows the evolution of classification results considering the complete SNPCC spectroscopic sample as a starting point.  Here we can clearly see the effect of the evolving sample sizes: accuracy and efficiency results oscillate, while purity and figure of merit remain indifferent to the learning strategy,  during the build-up phase (grey region). Once this phase is over, results start to differ and the AL with uncertainty sampling clearly surpasses the other two, achieving 80\% accuracy, 55\% purity and a figure of merit of 0.23, while the passive learning only goes up to 72\% accuracy, 45\% purity and figure of merit of 0.18. The canonical strategy continues to output better efficiency, but its loss in purity does not allow it to overcome even passive learning in figure of merit levels. 

\subsection{No initial training}
\label{subsec:notrain}

This leaves one open question: what should we do at the beginning of a given survey, when a training set with the same instrument characteristics (e.g. photometric system) is not yet available? Or even more drastically: if the algorithm is capable of building its own training sample, do we even need an initial training at all? The answer is no. 

Figure \ref{fig:TD2} shows how the classification results behave when the initial model is trained in 1 randomly selected object from the query sample, meaning we start with a random classifier. In this context, diagnostics  are meaningless until around 100 days (a little after the build-up phase) when all samples involved are under construction. After this period, AL with uncertainty sampling starts to dominate purity and, consequently, figure of merit results. After 150 observation days (or after 130 objects were added to the training), the active and passive learning strategies achieve purity levels comparable to the one obtained in the unrealistic full light curve scenario ($\sim 80\%$). Thus, at the end of the survey, AL efficiency results (27\%) are 80\% higher than the one obtained by passive learning (15\%), which translates into an almost doubled figure of merit (0.14 from AL and 0.08 from the canonical strategy). 
Compare these results with the \textit{initial state} of the full light curve analysis: figure \ref{fig:fullLC} (accuracy 60\%, efficiency 92\%, purity of 35\% and 0.15 figure of merit) was obtained using complete light curves for all objects, all SNe in the original SNPCC spectroscopic sample surviving the minimum number of epochs cuts (1094 objects) and the same random forest classifier. Final results of the real-time AL analysis (figure \ref{fig:TD2}) surpasses the full light curve initial state accuracy results in 33\%, more than doubles purity and achieves comparable figure of merit results. All of these while respecting the time evolution of observed epochs of only  161 SNe in the training set, or 15\% the number of objects in the original SNPCC spectroscopic sample. 

Accuracy levels of real time AL with (figure \ref{fig:TD1}) and without (figure \ref{fig:TD2}) the full initial training sample are comparable, while efficiency and figure of merit are higher for the former case. However, purity levels are 45\% higher without using the initial training. This is a natural consequence of the higher number of SNe Ia in the SNPCC spectroscopic sample (figure \ref{fig:samples}), which requires the algorithm to unlearn the preference for Ia classifications before it can achieve its full potential in purity results. Figure \ref{fig:TD2} also shows that regarding purity, passive learning is able to achieve the same results as those obtained with uncertainty sampling while efficiency is severely compromised --exactly the opposite behaviour shown by the canonical strategy. This is a consequence of the populations targeted by each of these strategies. By prioritizing brighter objects, the canonical strategy introduces a bias in the learning model towards SNIa classifications. On the other hand, by randomly sampling from the target, passive learning adds a larger number of non-Ia examples to the training, introducing an opposite bias, at least in the early stages of the survey.

In summary, given the intrinsic bias present in all canonically obtained samples, we advocate that the best strategy for a new survey is to construct its own training during the firsts running seasons. Letting its own photometric sample guide the decisions of spectroscopic targeting.  This is specially important if one has the final goal of supernova cosmology in mind, where the main objective is to maximize purity (minimize false positives) as well as many other scientific SN objectives.

\section{Batch-mode active learning}
\label{sec:semi}

In this section, we take another step towards a more realistic description of a spectroscopic follow-up scenario. Instead of choosing one SN at a time, spectroscopic follow-up resources for large scale surveys will probably allow a number of SNe to be spectroscopically observed per night. Thus, we need a strategy which allows us to extend the AL algorithm, optimizing our choice from one to a set (or a batch) of objects  at each iteration. We focus on two methods derived from the notion of uncertainty sampling: \textit{$N$-least certain} and \textit{Semi-supervised uncertainty sampling}. 

The $N$-least certain batch query strategy uses the same machinery described in the sequential uncertainty sampling method but, instead of choosing a single unlabelled example, it selects the $N$ objects with highest uncertainties, and queries all of them. This tactic carries a disadvantage, since  a set of objects whose predictions exhibit similar uncertainties will probably also be similar among themselves (i.e., will be close to each other in the feature space). Thus, querying for a set of labels is not likely to lead to a model much different than the one obtained by adding only the most uncertain object to the training set. In dealing with a batch mode scenario, we should also require that the elements of the batch be as diverse as possible (maximizing their distance in the feature space). 

Semi-supervised uncertainty sampling \citep[e.g.][]{hoi2008}, in contrast, avoids the need to call the oracle at each individual iteration by using the uncertainty associated to each predicted label as a proxy for class assignment. The algorithm must be trained in the available initial sample in order to create the first batch. 
The object with the greatest classification uncertainty is then identified.
Suppose this object has a probability $p$ of being SN Ia. A pseudo-label is then drawn from a Bernoulli distribution, where success is interpreted as ``Ia'' label (with probability $p$) and failure as ``non-Ia'' (with probability $1-p$). The object features and corresponding  pseudo-label are temporarily added to the training sample and the model is re-trained. This is repeated until we reach the size of the batch (see algorithm \ref{algo:semi}).
The benefit of using the model to produce  pseudo-labels comes with the inevitable uncertainty attached to model predictions: they come unwarranted. However, the problem attached to the $N$-least certain strategy is here, to a certain degree, overcome. Similar unlabelled instances are less likely to be included in the same batch. 

The optimum number of elements in each batch, $N$, is highly dependent on the particular combination of data set and classifier at hand. At each iteration, we are actually stretching the capabilities of the learning model in a feedback loop that cannot be expected to perform well for large batches. For the SNPCC data, our tests show that semi-supervised learning outperforms the N-least certain strategy for $N \in [2,8]$ with maximum results obtained with $N=5$.  

\begin{algorithm}
 \caption{Semi-supervised uncertainty sampling algorithm to identify which elements must be included in the batch.}
 \label{algo:semi}
 \KwData{Training set, $T_{tr}$, Unlabelled set, $T_u$ and batch size, N}
 \KwResult{List of data points to be queried} 
 R $\leftarrow$ [] \\
 Batch $\leftarrow$ [] \\
 \For{i=1,...,N}{
 Use $T_{tr}$ to train learning model $\mathcal{M}$ \\
 \For{$\mathbf{x} \in T_u$}{
     ID, class, Ia\_prob = $\mathcal{M}(\mathbf{x})$\\
     R $\leftarrow$ \{ID, class, Ia\_prob, $\mathbf{x}$\}
  }
  $r^* \leftarrow$ element in R with largest uncertainty\\
  Batch $\leftarrow r^*[1]$\\
  d $\sim \mathcal{B}(p=r^*[3])$\\
  \eIf{d $\equiv$ True}{
   y $\leftarrow r^*[2]$
   }{
   y $\leftarrow$ alternate class ($\neq r^*[2]$)
  }
  Add element to training set $T_{tr} \leftarrow$ ($r^*[4]$, y)\\
 }
 \KwRet Batch\\
\end{algorithm}

Figure \ref{fig:TD_batch5} shows classification results for canonical and passive learning (both at each iteration drawing 5 random elements from the pseudo-training and target sample respectively), AL via N-least certain and semi-supervised uncertainty sampling, when  the initial training consists of 5 randomly drawn objects from the query sample  and $N=5$. We see that in this scenario semi-supervised AL is able to achieve the same figure of merit   ($\sim 0.22$) as sequential uncertainty sampling when the entire initial training sample is available (figure \ref{fig:TD1}). However, it does so using only 63\%  of the number of objects for training (or 800 SNe in the training after 180 days, against 1263 SNe in the full training case). Moreover, although  efficiency results show  a steady increase until the end of the survey, purity achieves saturation levels ($\sim 0.8$ - the same as the final results obtained with the static full light curve scenario, figure \ref{fig:fullLC}) after only 100 days (corresponding to a training set with 405 objects). A numerical description of the final classification results and corresponding training size is shown in table~\ref{tab:samples}.

\begin{figure}
\includegraphics[width=\columnwidth]{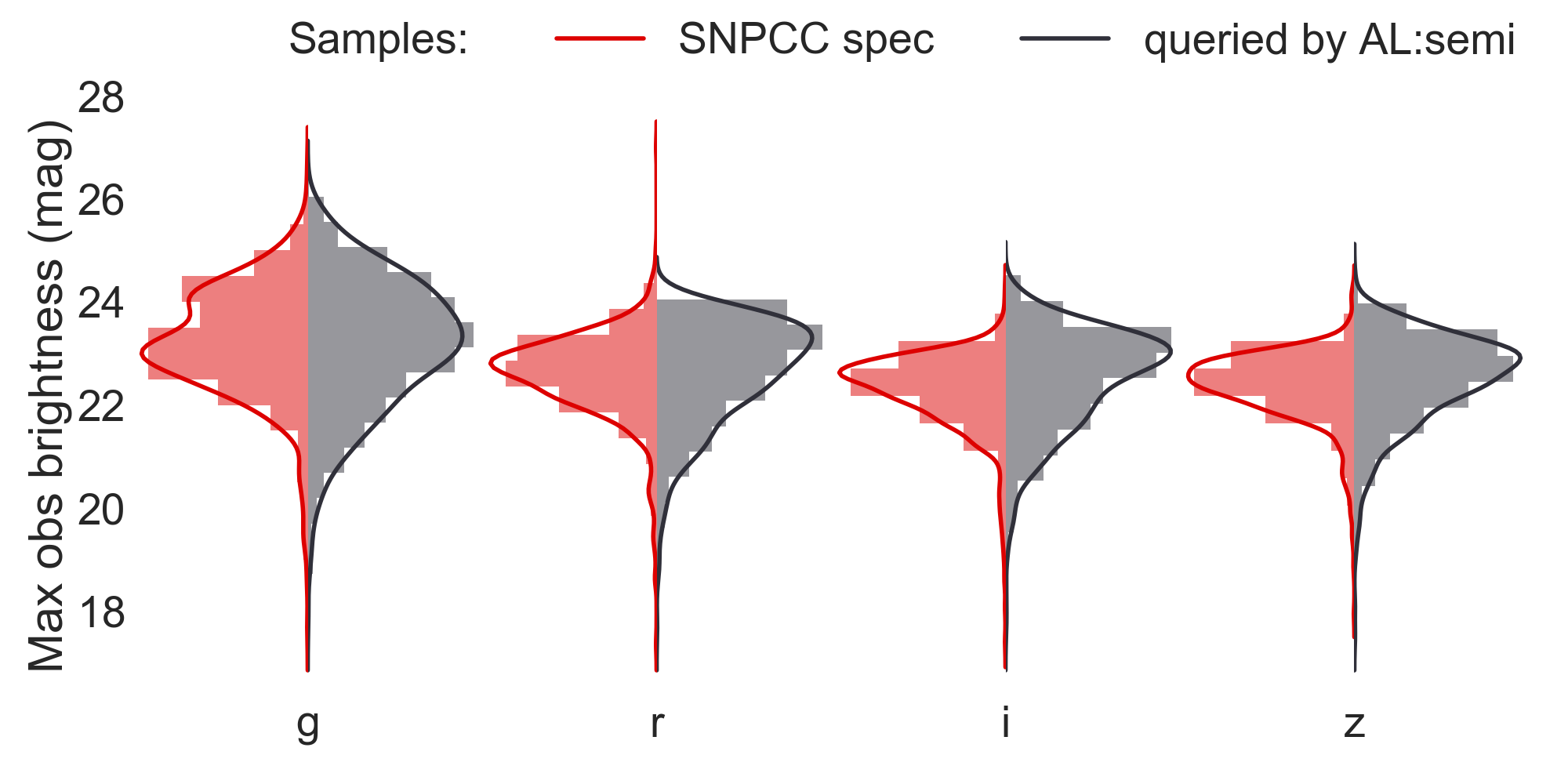}
\vspace{-0.5cm}
\includegraphics[width=\columnwidth]{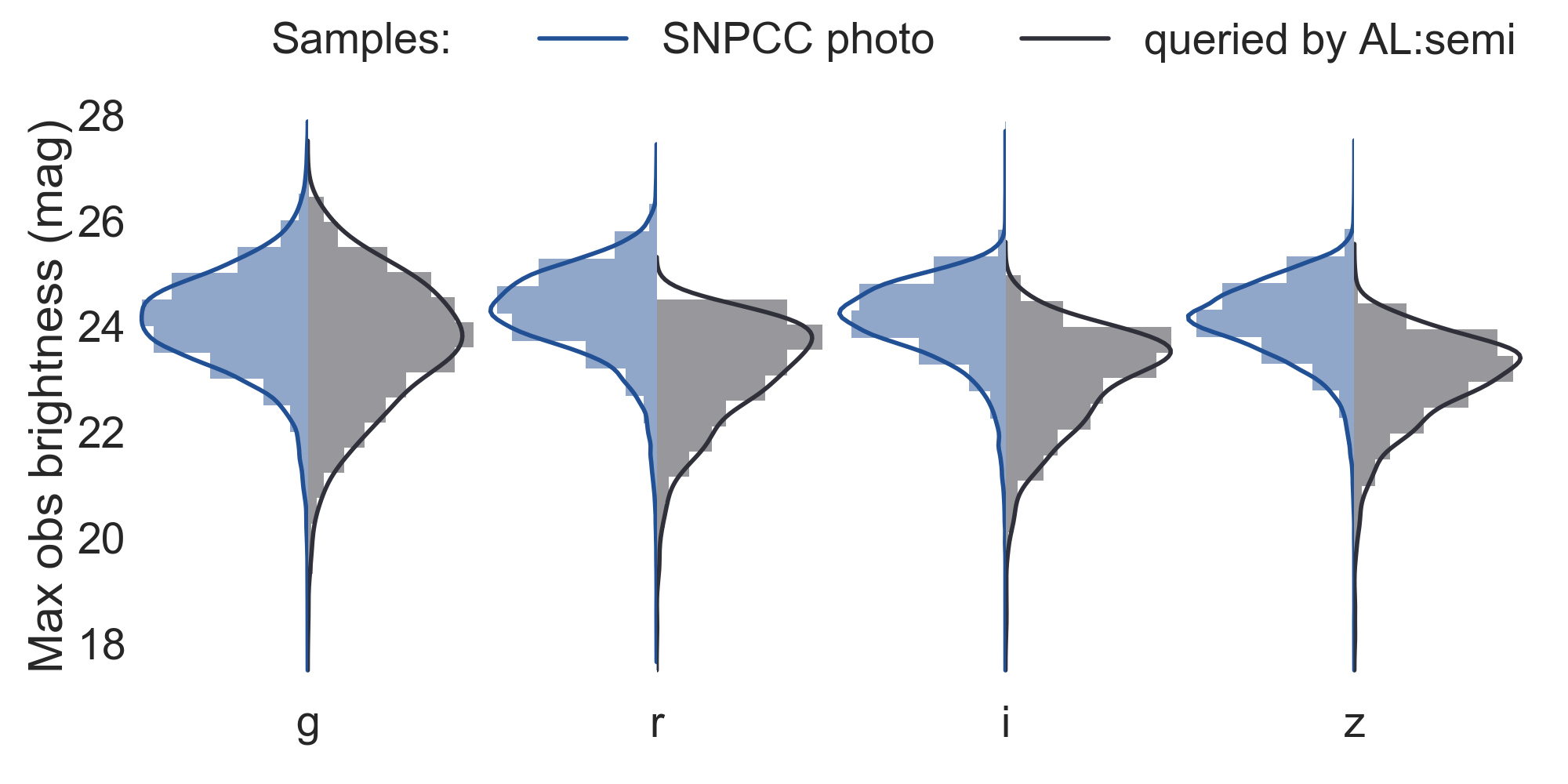}
\caption{Distributions of maximum observed brightness, in all DES filters, for the set of objects queried by AL via batch-mode semi-supervised uncertainty sampling with $N=5$ (dark grey). This is compared to distributions from SNPCC spectroscopic (red - top) and  SNPCC photometric (blue - bottom) samples.}
\label{fig:peakmag_samples}
\end{figure}

\begin{figure}
\includegraphics[width=\columnwidth]{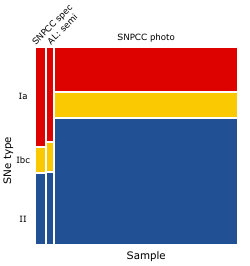}
\caption{Populations of different supernova types in the original SNPCC spectroscopic and photometric samples, and in time domain batch mode ($N=5$) semi-supervised AL query sample after 180 days of observations. The composition of the SNPCC samples are the same as shown in figure \ref{fig:types_ini}. The AL query sample holds 390 (48\%) Ia, 122 (15\%) Ibc and 298 (37\%) II.}
\label{fig:mondrian_AL}
\end{figure}

\begin{figure}
\includegraphics[width=\columnwidth]{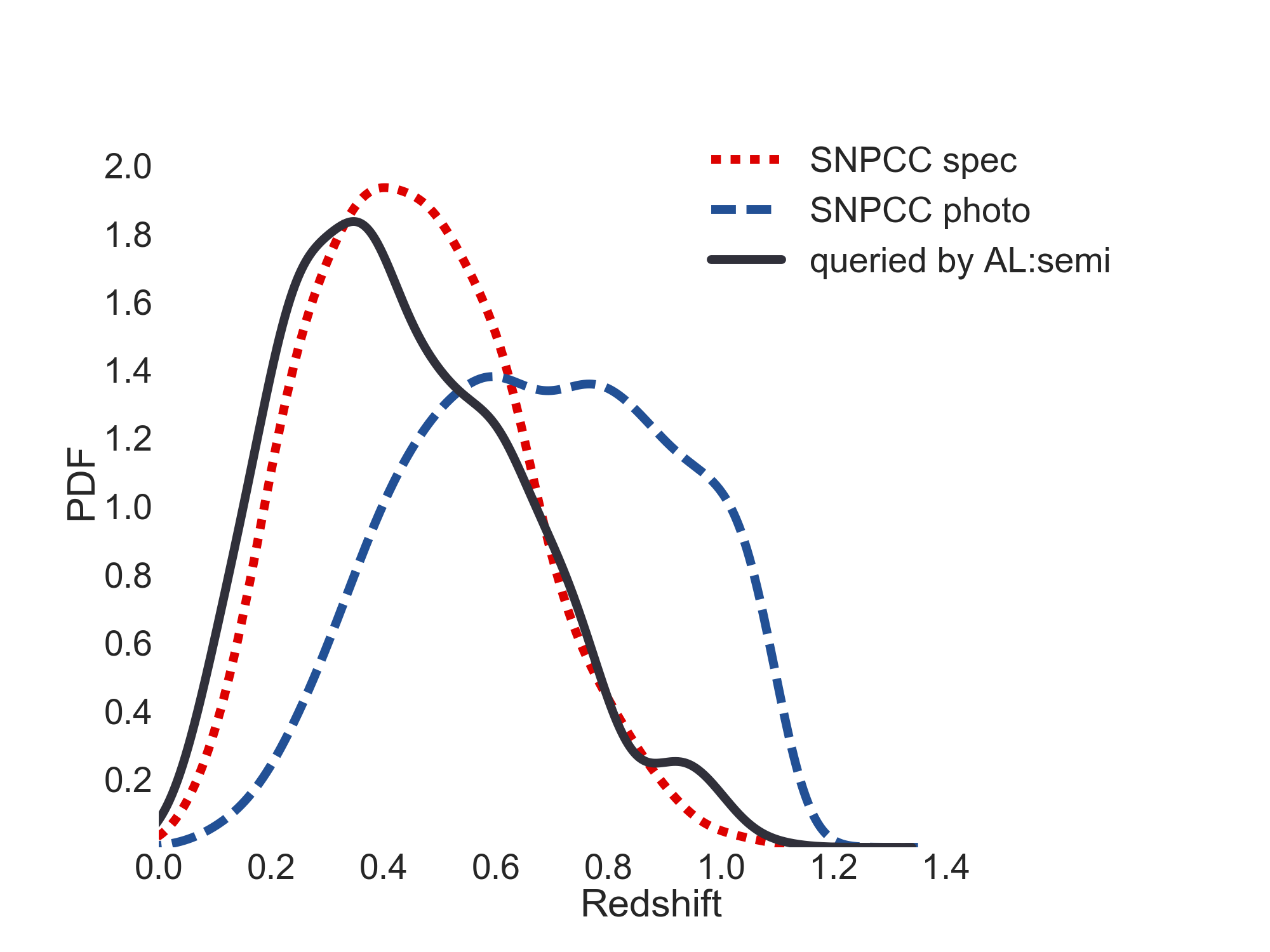}
\caption{Redshift distribution of the original SNPCC spectroscopic (red - dotted) and photometric (blue - dashed) samples, superimposed to the redshift distribution of the AL query set for the time domain semi-supervised batch mode AL strategy without the use of an initial training (dark blue - full). In each observation night, the algorithm queried for 5 SNe. The distribution shows redshift for the query sample after 180 observation nights.}. 
\label{fig:redshift_samples}
\end{figure}

From figure \ref{fig:TD_batch5} we see that samples containing the same number of objects lead to different classification results. Moreover, considering that the query sample only contains objects with $m_r\leq24$, we should not expect the set of objects queried by AL to be representative  of the target sample, despite the improvement in classification results driven by AL. 
This is clearly shown in figure \ref{fig:peakmag_samples}, where we compare distribution of maximum observed brightness  in each filter for the SNPCC spectroscopic (red) and photometric (blue) samples with the set of objects queried by AL  (dark grey). The latter provides a slight advantage in coverage when compared to the original spectroscopic sample, but it is still significantly different from the photometric distribution. A similar behaviour is found when we compare the populations of different SN types (figure \ref{fig:mondrian_AL}) and redshift distribution  (figure \ref{fig:redshift_samples}). These results confirm that, although a slight adjustment is necessary in order to optimize the allocation of spectroscopic time, a significant improvement in classification results may be achieved without a fully representative sample. 

\section{Telescope allocation time}
\label{sec:tel}

As a final remark, we must address the question of how much spectroscopic telescope time is required to obtain the labels queried by the AL algorithm -- in comparison to the time necessary to get all labels from SNPCC spectroscopic sample (canonical strategy).

In the realistic case of a survey adoption of the framework proposed here, a term taking into account the telescope time needed for spectroscopy observations must be added to the cost function of the AL algorithm. This was not explicitly taken into account in this paper, but we considered a constraint on magnitudes for the set of SNe available for spectroscopic follow-up ($r_{mag} \le 24$). We were able to estimate the integration time required for each object to achieve a given SNR by considering its magnitude and typical values for statistical noise of the sources\footnote{Namely, counts in the sky,  $\approx 13.8$ e$^{-}/s/pix$ and read-out noise, $\approx$ 8 e$^{-}$\citep[e.g.][]{Bolte15}.}. In the SNPCC spectroscopic sample, we considered the spectra taken at maximum brightness. For the set of AL queried objects, we used the magnitude at the epoch in which the object was queried. Considering a SNR of 10 (more than enough to enable classification) the ratio between the total spectroscopic time needed to get the labels for the SNPCC spectroscopic sample and the set of objects queried by semi-supervised AL  is $0.9992$. This indicates that the set of objects queried by AL would require less than $2.9$s more time than the SNPCC spectroscopic sample to be observed at each hour.
Also, if a more realistic estimation had been performed considering instrumental overheads, the set of objects queried by AL would have significant advantage, as it contains 26\% less objects than the SNPCC spectroscopic sample. This gives us the first indication that  AL-like approaches are feasible alternatives to minimize  instrumental usage and, at the same time, optimize scientific outcome of photometrically classified samples. 

For the specific case studied here, the high purity values achieved in early stages of the batch-mode AL,  accompanied by the steady increase in efficiency (figure \ref{fig:TD_batch5}) renders our final SN Ia sample optimally suited for photometric classification in cosmological analysis --albeit being smaller in number of objects and requiring almost the same amount of spectroscopic resources to be secured.

\begin{table*}
\begin{tabular}{c|cccc}
 & static, full LC & time domain  & time domain & \multirow{2}{*}{time domain} \\
 & initial training & initial training & initial training & \\
 & UNC & UNC & BATCH 5 & BATCH 5\\ 
 \hline
 training & \multirow{2}{*}{2093} & \multirow{2}{*}{1255}& \multirow{2}{*}{1093} & \multirow{2}{*}{810}\\
 size & & & & \\
 \hline
 accuracy & 0.89 & 0.80 & 0.85 & 0.83\\
 efficiency & 0.73& 0.78& 0.69 & 0.44\\
 purity & 0.78& 0.55& 0.69 & 0.76\\
 figure of merit & 0.39 & 0.23 & 0.31 & 0.22\\
\end{tabular}
\caption{Classification results for the AL by uncertainty sampling (UNC) and semi-supervised batch mode (BATCH 5) strategies. }
\label{tab:samples}
\end{table*}
\section{Conclusions}
\label{sec:conclusions}

Wide-field sky rolling surveys  will detect an unprecedented number of astronomical transients every night. However, the usefulness of these photometric data for cosmological analysis is conditioned on our ability to perform automatic and reliable light-curve classification using a very limited number of spectroscopic observations for validation. 
Traditional attempts to address this issue via supervised learning methods focus on a learning model that postulates a static, fully observed pair of spectroscopic (training) and photometric (target) samples. Such studies are paramount to assess the requirements and performance of different classifiers. Nevertheless, they fail to address fundamental aspects of astronomical data acquisition, which renders the problem ill-suited for text-book machine learning algorithms. The most crucial of these issues is the non-representativeness between spectroscopic and photometric samples. 

This mismatch has its origins in a follow-up strategy designed to maximize the number of spectroscopically confirmed SNe Ia for cosmology, resulting in a highly biased spectroscopic set --- and a sub-optimal training sample. Given such data configuration, not even the most suitable classifier can be expected to achieve its full potential. In this work, we advocate that any attempt to improve SN photometric classification must include a detailed strategy for constructing a more representative training set, without ignoring the constraints intrinsic to the observational process. 

Our proposed framework updates on a daily basis crucial steps of the SN photometric classification pipeline and uses active learning to optimize the scientific outcome from machine learning algorithms. On each day, we consider only the set of available observed epochs, and perform feature extraction via a parametric light-curve representation. The  identification of SNe available for spectroscopic targeting (objects with $m_{r}\leq 24$ on that day)  as a separate group from the full photometric sample is also updated daily. Finally, by using Active Learning (AL), we allow the algorithm itself to target those objects available for spectroscopic targeting that would maximally improve the learning model if added to the training set. Using the proposed semi-supervised batch-mode AL, we designate an optimal
set of new objects to be spectroscopically observed on each night. Once the batch is identified, the model is re-trained and new spectroscopic targets are selected for the subsequent night. This method avoids the necessity of an initial training sample: it starts with a random classifier, allowing the algorithm to construct an optimal training sample from scratch, specifically adapted  to the survey at hand. The framework was successfully applied to the simulated data released after the \textit{Supernova Photometric Classification Challenge} \citep[SNPCC][]{kessler2010}.

Our results show that the proposed framework is able to achieve high purity levels, $\sim 80\%$,  after only 100 observation days --- which corresponds to a training set of only 400 objects. After 180 observation days, or 800 queries, we are able to reach a figure of merit of 0.22 (figure \ref{fig:TD_batch5}) --- highly above the values we would have obtained by using the canonical strategy for the idealized full training and full light curve analysis (which achieves $35\%$ purity and figure of merit of 0.14, red curve in figure \ref{fig:fullLC}).

After showing the classification improvements achieved with the AL algorithm, we examined the characteristics of the set of objects queried by AL. As expected, the   $m_{r}\leq 24$ requirement does not allow this set to deviate strongly from the characteristics of the SNPCC spectroscopic sample (see figures \ref{fig:peakmag_samples}, \ref{fig:redshift_samples} and \ref{fig:mondrian_AL}). However, these small differences translate into significant improvements in purity, and consequently, figure of merit results. 
This comes with no penalty whatsoever in the necessary observational time for spectroscopic follow-up. The ratio of integration times required to observe  the complete set of queried objects and the SNPCC spectroscopic sample is close to unity. This is still additional evidence that Active Learning is a viable strategy to optimize the distribution of spectroscopic resources.

In all our results, we observe that the canonical targeting strategy, which prioritizes the spectroscopic follow-up of bright events which resemble SNe Ia, has higher efficiency results (which means that this strategy is more successful in identifying a high number of SNe Ia in the target sample). However, as the diversity present in the training sample is very low, purity levels are always compromised. The canonical strategy is successful in targeting SNe Ia, but it is not optimal when the goal is to construct a training sample for machine learning classifiers. 

In the present analysis, we restricted ourselves to the SNe photometric classification case of SNe Ia versus non-Ia. However, we stress that our methodology is easily adaptable to more general classifications that include other transients \citep[e.g.][]{narayan2018}. This exercise will be especially informative when applied to the data from the upcoming \textit{Photometric LSST Astronomical Time-series Classification Challenge}\footnote{\href{https://plasticcblog.wordpress.com/}{https://plasticcblog.wordpress.com/}} (PLAsTiCC).

Finally, it is reasonable to expect that different combinations of classifiers and feature extraction methods will react differently to the iterative AL process. The same is valid for the AL algorithm itself. Moreover, there are other practical issues that we did not take into account, like the time delay necessary to treat the spectra and provide a classification or the fact that sometimes one spectra is not enough to get a classification. In summary, we recognize that each item in our pipeline can be refined, potentially leading to even more drastic improvements in classification results. The results we show here are the first evidence that an iterative learning process adapted to the specificities of astronomical observations can lead to significant optimization in the allocation of observational resources. 

Adapting to the era of big data in astronomy will entail adapting machine learning techniques to the unique reality of astronomical observations. This requires tackling fundamental issues which will always be present in astronomical data (e.g. the discrepancy between training and test samples and the time evolution of a transient survey).  Astronomy has once again the opportunity to provide ground for developments in other research areas by providing unique data situations not commonly present in other scenarios, as long as a consistent interdisciplinary environment is available. We are convinced that the most exciting part of this endeavour is still to come.

\section*{Acknowledgements}
This work was created during the $\rm 4^{th}$ COIN Residence Program\footnote{\href{https://iaacoin.wixsite.com/crp2017}{https://iaacoin.wixsite.com/crp2017}} (CRP\#4), held in Clermont-Ferrand, France on August 2017, with support from  Universit\'e Clermont-Auvergne and  La R\'egion Auvergne-Rh\^one-Alpes. This project is financially supported by CNRS as part of its MOMENTUM programme over the 2018-2020 period. EEOI thanks Michele Sasdelli for comments on the draft and Isobel Hook for useful  discussions. AKM acknowledges the support from the Portuguese Funda\c c\~ao para a Ci\^encia e a Tecnologia (FCT) through grants SFRH/BPD/74697/2010, from the Portuguese Strategic Programme UID/FIS/00099/2013 for CENTRA, the ESA contract AO/1-7836/14/NL/HB and Caltech Division of Physics, Mathematics and Astronomy for hosting a research leave during 2017-2018, when this paper was prepared. 
RSS thanks the support from NASA under the Astrophysics Theory 
Program Grant 14-ATP14-0007. RB acknowledges support from the National Science Foundation (NSF) award 1616974 and the NKFI NN 114560 grant of Hungary. BQ acknowledges financial support from CNPq-Brazil under the process number 205459/2014-5. AZV acknowledges financial support from CNPq. AM thanks partial support from NSF through grants AST-0909182, AST-1313422, AST-1413600, and AST-1518308.
This work has made use of the computing facilities of the Laboratory of Astroinformatics (IAG/USP, NAT/Unicsul), whose purchase was made possible by the Brazilian agency FAPESP (grant 2009/54006-4) and the INCT-A.
This work was partly supported by the Center for Advanced Computing and Data Systems (CACDS), and by the Texas Institute for Measurement, Evaluation, and Statistics (TIMES) at the University of Houston.
This project has been supported by a Marie Sklodowska-Curie Innovative Training Network Fellowship of the European Commission's Horizon 2020 Programme under contract number 675440 AMVA4NewPhysics.

The Cosmostatistics  Initiative\footnote{\href{https://github.com/COINtoolbox}{https://github.com/COINtoolbox}} (COIN) is a non-profit organization whose aim is to nourish the synergy between astrophysics, cosmology, statistics, and machine learning communities.
 This work benefited from the following collaborative platforms: \texttt{Overleaf}\footnote{\url{https://www.overleaf.com}}, \texttt{Github}\footnote{\url{https://github.com}}, and \texttt{Slack}\footnote{\url{https://slack.com/}}.
 
\bibliographystyle{mnras}
\bibliography{ref}

\begin{thebibliography}{}
\makeatletter
\relax
\def\mn@urlcharsother{\let\do\@makeother \do\$\do\&\do\#\do\^\do\_\do\%\do\~}
\def\mn@doi{\begingroup\mn@urlcharsother \@ifnextchar [ {\mn@doi@}
  {\mn@doi@[]}}
\def\mn@doi@[#1]#2{\def\@tempa{#1}\ifx\@tempa\@empty \href
  {http://dx.doi.org/#2} {doi:#2}\else \href {http://dx.doi.org/#2} {#1}\fi
  \endgroup}
\def\mn@eprint#1#2{\mn@eprint@#1:#2::\@nil}
\def\mn@eprint@arXiv#1{\href {http://arxiv.org/abs/#1} {{\tt arXiv:#1}}}
\def\mn@eprint@dblp#1{\href {http://dblp.uni-trier.de/rec/bibtex/#1.xml}
  {dblp:#1}}
\def\mn@eprint@#1:#2:#3:#4\@nil{\def\@tempa {#1}\def\@tempb {#2}\def\@tempc
  {#3}\ifx \@tempc \@empty \let \@tempc \@tempb \let \@tempb \@tempa \fi \ifx
  \@tempb \@empty \def\@tempb {arXiv}\fi \@ifundefined
  {mn@eprint@\@tempb}{\@tempb:\@tempc}{\expandafter \expandafter \csname
  mn@eprint@\@tempb\endcsname \expandafter{\@tempc}}}

\bibitem[\protect\citeauthoryear{Balcan, Beygelzimer  \& Langford}{Balcan
  et~al.}{2009}]{Balcan09}
Balcan M.~F.,  Beygelzimer A.,   Langford J.,  2009, Journal of Computer and
  System Sciences, 75, 78

\bibitem[\protect\citeauthoryear{{Bazin} et~al.}{{Bazin}
  et~al.}{2009}]{Bazin09}
{Bazin} G.,  et~al., 2009, \mn@doi [\aap] {10.1051/0004-6361/200911847}, 499,
  653

\bibitem[\protect\citeauthoryear{{Betoule} et~al.,}{{Betoule}
  et~al.}{2014}]{betoule2014}
{Betoule} M.,  et~al., 2014, \mn@doi [\aap] {10.1051/0004-6361/201423413},
  \href {http://cdsads.u-strasbg.fr/abs/2014A%26A...568A..22B} {568, A22}

\bibitem[\protect\citeauthoryear{{Bolte}}{{Bolte}}{2015}]{Bolte15}
{Bolte} M.,  2015, Modern Observational Techniques, \url
  {http://www.ucolick.org/~bolte/AY257/s_n.pdf}

\bibitem[\protect\citeauthoryear{Breiman}{Breiman}{2001}]{Breiman01}
Breiman L.,  2001, Machine Learning, 45, 5

\bibitem[\protect\citeauthoryear{Breiman, Friedman, Olshen  \& Stone}{Breiman
  et~al.}{1984}]{breiman1984}
Breiman L.,  Friedman J.~H.,  Olshen R.~A.,   Stone C.~J.,  1984,
  Classification and Regression Trees.
Wadsworth and Brooks, Monterey, CA

\bibitem[\protect\citeauthoryear{{Campbell} et~al.}{{Campbell}
  et~al.}{2013}]{Campbell13}
{Campbell} H.,  et~al., 2013, \mn@doi [\apj] {10.1088/0004-637X/763/2/88},
  \href {http://adsabs.harvard.edu/abs/2013ApJ...763...88C} {763, 88}

\bibitem[\protect\citeauthoryear{{Charnock} \& {Moss}}{{Charnock} \&
  {Moss}}{2017}]{charnock2017}
{Charnock} T.,  {Moss} A.,  2017, \mn@doi [\apjl] {10.3847/2041-8213/aa603d},
  \href {http://adsabs.harvard.edu/abs/2017ApJ...837L..28C} {837, L28}

\bibitem[\protect\citeauthoryear{{Childress} et~al.}{{Childress}
  et~al.}{2017}]{Childress17}
{Childress} M.~J.,  et~al., 2017, preprint, \href
  {http://adsabs.harvard.edu/abs/2017arXiv170804526C} {} (\mn@eprint {arXiv}
  {1708.04526})

\bibitem[\protect\citeauthoryear{Cohn, Ghahramani  \& Jordan}{Cohn
  et~al.}{1996}]{Cohn96}
Cohn D.~A.,  Ghahramani Z.,   Jordan M.~I.,  1996, Journal of Artificial
  Intelligence Research, 4, 129

\bibitem[\protect\citeauthoryear{{Conley} et~al.}{{Conley}
  et~al.}{2011}]{Conley11}
{Conley} A.,  et~al., 2011, \mn@doi [\apjs] {10.1088/0067-0049/192/1/1}, \href
  {http://adsabs.harvard.edu/abs/2011ApJS..192....1C} {192, 1}

\bibitem[\protect\citeauthoryear{Cover \& Thomas}{Cover \&
  Thomas}{2006}]{Cover06}
Cover T.~M.,  Thomas J.~A.,  2006, Elements of Information Theory (Wiley Series
  in Telecommunications and Signal Processing).
Wiley-Interscience

\bibitem[\protect\citeauthoryear{{Dai}, {Kuhlmann}, {Wang}  \& {Kovacs}}{{Dai}
  et~al.}{2017}]{dai2017}
{Dai} M.,  {Kuhlmann} S.,  {Wang} Y.,   {Kovacs} E.,  2017, preprint, \href
  {http://adsabs.harvard.edu/abs/2017arXiv170105689D} {} (\mn@eprint {arXiv}
  {1701.05689})

\bibitem[\protect\citeauthoryear{DeBarr \& Wechsler}{DeBarr \&
  Wechsler}{2009}]{debarr2009}
DeBarr D.,  Wechsler H.,  2009, in Sixth Conference on Email and Anti-Spam.
  Mountain View, California. pp~1--6

\bibitem[\protect\citeauthoryear{{Foley} \& {Mandel}}{{Foley} \&
  {Mandel}}{2013}]{Foley13}
{Foley} R.~J.,  {Mandel} K.,  2013, \mn@doi [\apj]
  {10.1088/0004-637X/778/2/167}, \href
  {http://adsabs.harvard.edu/abs/2013ApJ...778..167F} {778, 167}

\bibitem[\protect\citeauthoryear{{Gamow}}{{Gamow}}{1948}]{Gamow1948}
{Gamow} G.,  1948, \mn@doi [\nat] {10.1038/162680a0}, \href
  {http://adsabs.harvard.edu/abs/1948Natur.162..680G} {162, 680}

\bibitem[\protect\citeauthoryear{{Goobar} \& {Leibundgut}}{{Goobar} \&
  {Leibundgut}}{2011}]{goobar2011}
{Goobar} A.,  {Leibundgut} B.,  2011, \mn@doi [Annual Review of Nuclear and
  Particle Science] {10.1146/annurev-nucl-102010-130434}, \href
  {http://adsabs.harvard.edu/abs/2011ARNPS..61..251G} {61, 251}

\bibitem[\protect\citeauthoryear{Gupta, Pampana, Vilalta, Ishida  \& de
  Souza}{Gupta et~al.}{2016}]{dhargupta17}
Gupta K.~D.,  Pampana R.,  Vilalta R.,  Ishida E. E.~O.,   de Souza R.~S.,
  2016, in 2016 IEEE Symposium Series on Computational Intelligence (SSCI).

\bibitem[\protect\citeauthoryear{Hillebrandt \& Niemeyer}{Hillebrandt \&
  Niemeyer}{2000}]{hillebrandt2000}
Hillebrandt W.,  Niemeyer J.~C.,  2000, \mn@doi [Annual Review of Astronomy and
  Astrophysics] {10.1146/annurev.astro.38.1.191}, 38, 191

\bibitem[\protect\citeauthoryear{{Hlozek} et~al.,}{{Hlozek}
  et~al.}{2012}]{hlozek2012}
{Hlozek} R.,  et~al., 2012, \mn@doi [\apj] {10.1088/0004-637X/752/2/79}, \href
  {http://adsabs.harvard.edu/abs/2012ApJ...752...79H} {752, 79}

\bibitem[\protect\citeauthoryear{Hoi, Jin, Zhu  \& Lyu}{Hoi
  et~al.}{2008}]{hoi2008}
Hoi S. C.~H.,  Jin R.,  Zhu J.,   Lyu M.~R.,  2008, in 2008 IEEE Conference on
  Computer Vision and Pattern Recognition. pp~1--7,
  \mn@doi{10.1109/CVPR.2008.4587350}

\bibitem[\protect\citeauthoryear{{Hoyle}, {Paech}, {Rau}, {Seitz}  \&
  {Weller}}{{Hoyle} et~al.}{2016}]{hoyle2016}
{Hoyle} B.,  {Paech} K.,  {Rau} M.~M.,  {Seitz} S.,   {Weller} J.,  2016,
  \mn@doi [\mnras] {10.1093/mnras/stw563}, \href
  {http://adsabs.harvard.edu/abs/2016MNRAS.458.4498H} {458, 4498}

\bibitem[\protect\citeauthoryear{{Ishida} \& {de Souza}}{{Ishida} \& {de
  Souza}}{2013}]{ishida2013}
{Ishida} E.~E.~O.,  {de Souza} R.~S.,  2013, \mn@doi [\mnras]
  {10.1093/mnras/sts650}, \href
  {http://adsabs.harvard.edu/abs/2013MNRAS.430..509I} {430, 509}

\bibitem[\protect\citeauthoryear{{Johnson} \& {Crotts}}{{Johnson} \&
  {Crotts}}{2006}]{Johnson06}
{Johnson} B.~D.,  {Crotts} A.~P.~S.,  2006, \mn@doi [\aj] {10.1086/503528},
  \href {http://adsabs.harvard.edu/abs/2006AJ....132..756J} {132, 756}

\bibitem[\protect\citeauthoryear{{Jones} et~al.,}{{Jones}
  et~al.}{2017}]{jones2017}
{Jones} D.~O.,  et~al., 2017, \mn@doi [\apj] {10.3847/1538-4357/aa767b}, \href
  {http://adsabs.harvard.edu/abs/2017ApJ...843....6J} {843, 6}

\bibitem[\protect\citeauthoryear{{Karpenka}, {Feroz}  \& {Hobson}}{{Karpenka}
  et~al.}{2013}]{karpenka2013}
{Karpenka} N.~V.,  {Feroz} F.,   {Hobson} M.~P.,  2013, \mn@doi [\mnras]
  {10.1093/mnras/sts412}, \href
  {http://adsabs.harvard.edu/abs/2013MNRAS.429.1278K} {429, 1278}

\bibitem[\protect\citeauthoryear{{Kessler} et~al.,}{{Kessler}
  et~al.}{2010}]{kessler2010}
{Kessler} R.,  et~al., 2010, \mn@doi [Publications of the Astronomical Society
  of Pacific] {10.1086/657607}, \href
  {http://adsabs.harvard.edu/abs/2010PASP..122.1415K} {122, 1415}

\bibitem[\protect\citeauthoryear{Kranjc, Smailovi{\'c}, Podpe{\v{c}}an,
  Gr{\v{c}}ar, {\v{Z}}nidar{\v{s}}i{\v{c}}  \& Lavra{\v{c}}}{Kranjc
  et~al.}{2015}]{kranjc2015}
Kranjc J.,  Smailovi{\'c} J.,  Podpe{\v{c}}an V.,  Gr{\v{c}}ar M.,
  {\v{Z}}nidar{\v{s}}i{\v{c}} M.,   Lavra{\v{c}} N.,  2015, Information
  Processing \& Management, 51, 187

\bibitem[\protect\citeauthoryear{{Kuznetsova} \& {Connolly}}{{Kuznetsova} \&
  {Connolly}}{2007}]{Kuznetsova07}
{Kuznetsova} N.~V.,  {Connolly} B.~M.,  2007, \mn@doi [\apj] {10.1086/511814},
  \href {http://adsabs.harvard.edu/abs/2007ApJ...659..530K} {659, 530}

\bibitem[\protect\citeauthoryear{Liu}{Liu}{2004}]{liu2004}
Liu Y.,  2004, Journal of chemical information and computer sciences, 44, 1936

\bibitem[\protect\citeauthoryear{{Lochner}, {McEwen}, {Peiris}, {Lahav}  \&
  {Winter}}{{Lochner} et~al.}{2016}]{lochner2016}
{Lochner} M.,  {McEwen} J.~D.,  {Peiris} H.~V.,  {Lahav} O.,   {Winter} M.~K.,
  2016, \mn@doi [\apjs] {10.3847/0067-0049/225/2/31}, \href
  {http://adsabs.harvard.edu/abs/2016ApJS..225...31L} {225, 31}

\bibitem[\protect\citeauthoryear{Madsen, Nielsen  \& Tingleff}{Madsen
  et~al.}{2004}]{LevenbergMarquardt}
Madsen K.,  Nielsen H.~B.,   Tingleff O.,  2004, Methods for Non-Linear Least
  Squares Problems (2nd ed.)

\bibitem[\protect\citeauthoryear{{Masters} et~al.,}{{Masters}
  et~al.}{2015}]{Masters2015}
{Masters} D.,  et~al., 2015, \mn@doi [\apj] {10.1088/0004-637X/813/1/53}, \href
  {http://adsabs.harvard.edu/abs/2015ApJ...813...53M} {813, 53}

\bibitem[\protect\citeauthoryear{{M{\"o}ller} et~al.,}{{M{\"o}ller}
  et~al.}{2016}]{moller2016}
{M{\"o}ller} A.,  et~al., 2016, \mn@doi [\jcap]
  {10.1088/1475-7516/2016/12/008}, \href
  {http://adsabs.harvard.edu/abs/2016JCAP...12..008M} {12, 008}

\bibitem[\protect\citeauthoryear{{Narayan} et~al.,}{{Narayan}
  et~al.}{2018}]{narayan2018}
{Narayan} G.,  et~al., 2018, preprint, \href
  {http://adsabs.harvard.edu/abs/2018arXiv180107323N} {} (\mn@eprint {arXiv}
  {1801.07323})

\bibitem[\protect\citeauthoryear{{Naul}, {Bloom}, {P{\'e}rez}  \& {van der
  Walt}}{{Naul} et~al.}{2018}]{Naul2018}
{Naul} B.,  {Bloom} J.~S.,  {P{\'e}rez} F.,   {van der Walt} S.,  2018, \mn@doi
  [Nature Astronomy] {10.1038/s41550-017-0321-z}, \href
  {http://adsabs.harvard.edu/abs/2018NatAs...2..151N} {2, 151}

\bibitem[\protect\citeauthoryear{{Newling} et~al.,}{{Newling}
  et~al.}{2011}]{newling2011}
{Newling} J.,  et~al., 2011, \mn@doi [\mnras]
  {10.1111/j.1365-2966.2011.18514.x}, \href
  {http://adsabs.harvard.edu/abs/2011MNRAS.414.1987N} {414, 1987}

\bibitem[\protect\citeauthoryear{{Perlmutter} et~al.,}{{Perlmutter}
  et~al.}{1999}]{perlmutter1999}
{Perlmutter} S.,  et~al., 1999, \mn@doi [\apj] {10.1086/307221}, \href
  {http://adsabs.harvard.edu/abs/1999ApJ...517..565P} {517, 565}

\bibitem[\protect\citeauthoryear{{Perrett} et~al.}{{Perrett}
  et~al.}{2010}]{Perrett10}
{Perrett} K.,  et~al., 2010, \mn@doi [\aj] {10.1088/0004-6256/140/2/518}, \href
  {http://adsabs.harvard.edu/abs/2010AJ....140..518P} {140, 518}

\bibitem[\protect\citeauthoryear{{Phillips}}{{Phillips}}{1993}]{Phillips93}
{Phillips} M.~M.,  1993, \apjl, 413, L105

\bibitem[\protect\citeauthoryear{{Planck Collaboration} et~al.,}{{Planck
  Collaboration} et~al.}{2016}]{Planck2016}
{Planck Collaboration} et~al., 2016, \mn@doi [\aap]
  {10.1051/0004-6361/201527101}, \href
  {http://adsabs.harvard.edu/abs/2016A%26A...594A...1P} {594, A1}

\bibitem[\protect\citeauthoryear{{Poznanski}, {Gal-Yam}, {Maoz}, {Filippenko},
  {Leonard}  \& {Matheson}}{{Poznanski} et~al.}{2002}]{Poznanski02}
{Poznanski} D.,  {Gal-Yam} A.,  {Maoz} D.,  {Filippenko} A.~V.,  {Leonard}
  D.~C.,   {Matheson} T.,  2002, \mn@doi [\pasp] {10.1086/341741}, \href
  {http://adsabs.harvard.edu/abs/2002PASP..114..833P} {114, 833}

\bibitem[\protect\citeauthoryear{{Poznanski}, {Maoz}  \& {Gal-Yam}}{{Poznanski}
  et~al.}{2007}]{Poznanski07}
{Poznanski} D.,  {Maoz} D.,   {Gal-Yam} A.,  2007, \mn@doi [\aj]
  {10.1086/520956}, \href {http://adsabs.harvard.edu/abs/2007AJ....134.1285P}
  {134, 1285}

\bibitem[\protect\citeauthoryear{{Revsbech}, {Trotta}  \& {van Dyk}}{{Revsbech}
  et~al.}{2017}]{revsbech2017}
{Revsbech} E.~A.,  {Trotta} R.,   {van Dyk} D.~A.,  2017, preprint, \href
  {http://adsabs.harvard.edu/abs/2017arXiv170603811R} {} (\mn@eprint {arXiv}
  {1706.03811})

\bibitem[\protect\citeauthoryear{{Richards}, {Homrighausen}, {Freeman},
  {Schafer}  \& {Poznanski}}{{Richards} et~al.}{2012a}]{richards2012}
{Richards} J.~W.,  {Homrighausen} D.,  {Freeman} P.~E.,  {Schafer} C.~M.,
  {Poznanski} D.,  2012a, \mn@doi [\mnras] {10.1111/j.1365-2966.2011.19768.x},
  \href {http://adsabs.harvard.edu/abs/2012MNRAS.419.1121R} {419, 1121}

\bibitem[\protect\citeauthoryear{{Richards} et~al.,}{{Richards}
  et~al.}{2012b}]{richards2012b}
{Richards} J.~W.,  et~al., 2012b, \mn@doi [\apj] {10.1088/0004-637X/744/2/192},
  \href {http://adsabs.harvard.edu/abs/2012ApJ...744..192R} {744, 192}

\bibitem[\protect\citeauthoryear{{Riess} et~al.,}{{Riess}
  et~al.}{1998}]{riess1998}
{Riess} A.~G.,  et~al., 1998, \mn@doi [\aj] {10.1086/300499}, \href
  {http://adsabs.harvard.edu/abs/1998AJ....116.1009R} {116, 1009}

\bibitem[\protect\citeauthoryear{{Rodney} \& {Tonry}}{{Rodney} \&
  {Tonry}}{2009}]{Rodney09}
{Rodney} S.~A.,  {Tonry} J.~L.,  2009, \mn@doi [\apj]
  {10.1088/0004-637X/707/2/1064}, \href
  {http://adsabs.harvard.edu/abs/2009ApJ...707.1064R} {707, 1064}

\bibitem[\protect\citeauthoryear{{Sako} et~al.,}{{Sako}
  et~al.}{2008}]{sako2008}
{Sako} M.,  et~al., 2008, \mn@doi [\aj] {10.1088/0004-6256/135/1/348}, \href
  {http://adsabs.harvard.edu/abs/2008AJ....135..348S} {135, 348}

\bibitem[\protect\citeauthoryear{Settles}{Settles}{2012}]{Settles12}
Settles B.,  2012, Active Learning.
Morgan \& Claypool

\bibitem[\protect\citeauthoryear{{Solorio}, {Fuentes}, {Terlevich}  \&
  {Terlevich}}{{Solorio} et~al.}{2005}]{solorio2005}
{Solorio} T.,  {Fuentes} O.,  {Terlevich} R.,   {Terlevich} E.,  2005, \mn@doi
  [\mnras] {10.1111/j.1365-2966.2005.09456.x}, \href
  {http://adsabs.harvard.edu/abs/2005MNRAS.363..543S} {363, 543}

\bibitem[\protect\citeauthoryear{{Spergel} et~al.,}{{Spergel}
  et~al.}{2007}]{Spergel2007}
{Spergel} D.~N.,  et~al., 2007, \mn@doi [\apjs] {10.1086/513700}, \href
  {http://adsabs.harvard.edu/abs/2007ApJS..170..377S} {170, 377}

\bibitem[\protect\citeauthoryear{{Sullivan} et~al.}{{Sullivan}
  et~al.}{2006}]{Sullivan06}
{Sullivan} M.,  et~al., 2006, \mn@doi [\aj] {10.1086/499302}, 131, 960

\bibitem[\protect\citeauthoryear{Thompson, Califf  \& Mooney}{Thompson
  et~al.}{1999}]{thompson1999}
Thompson C.~A.,  Califf M.~E.,   Mooney R.~J.,  1999, in ICML. pp 406--414

\bibitem[\protect\citeauthoryear{{Tripp}}{{Tripp}}{1998}]{Tripp98}
{Tripp} R.,  1998, \aap, 331, 815

\bibitem[\protect\citeauthoryear{{Varughese}, {von Sachs}, {Stephanou}  \&
  {Bassett}}{{Varughese} et~al.}{2015}]{varughese2015}
{Varughese} M.~M.,  {von Sachs} R.,  {Stephanou} M.,   {Bassett} B.~A.,  2015,
  \mn@doi [\mnras] {10.1093/mnras/stv1816}, \href
  {http://adsabs.harvard.edu/abs/2015MNRAS.453.2848V} {453, 2848}

\bibitem[\protect\citeauthoryear{Vilalta, Ishida, Beck, Sutrisno, de Souza  \&
  Mahabal}{Vilalta et~al.}{2017}]{vilalta17}
Vilalta R.,  Ishida E. E.~O.,  Beck R.,  Sutrisno R.,  de Souza R.~S.,
  Mahabal A.,  2017, in 2017 IEEE Symposium Series on Computational
  Intelligence (SSCI).

\bibitem[\protect\citeauthoryear{{Wang}, {Gjergo}  \& {Kuhlmann}}{{Wang}
  et~al.}{2015}]{wang2015}
{Wang} Y.,  {Gjergo} E.,   {Kuhlmann} S.,  2015, \mn@doi [\mnras]
  {10.1093/mnras/stv1090}, \href
  {http://adsabs.harvard.edu/abs/2015MNRAS.451.1955W} {451, 1955}

\bibitem[\protect\citeauthoryear{Xia, Protopapas  \& Doshi-Velez}{Xia
  et~al.}{2016}]{xia2016}
Xia X.,  Protopapas P.,   Doshi-Velez F.,  2016, Cost-Sensitive Batch Mode
  Active Learning: Designing Astronomical Observation by Optimizing Telescope
  Time and Telescope Choice.
pp 477--485, \mn@doi{10.1137/1.9781611974348.54}

\bibitem[\protect\citeauthoryear{{Yang}, {Lee}, {Chung}, {Wu}, {Chen}  \&
  {Lin}}{{Yang} et~al.}{2017}]{yang2017}
{Yang} Y.-Y.,  {Lee} S.-C.,  {Chung} Y.-A.,  {Wu} T.-E.,  {Chen} S.-A.,   {Lin}
  H.-T.,  2017, preprint, \href
  {http://adsabs.harvard.edu/abs/2017arXiv171000379Y} {} (\mn@eprint {arXiv}
  {1710.00379})

\makeatother
\end{thebibliography}

\appendix
\section{Active Learning}
\label{ap:AL}

We now give a more detailed description of the Active Learning (AL) framework presented in section \ref{sec:AL}. As mentioned above, AL is an area of study in machine learning that suggests which are the most informative samples to add to the training set to enhance the current classification model \citep{Settles12,Balcan09,Cohn96}. 
AL assumes two sample sets: $T_{\rm tr}$ and $T_u$, where $T_{\rm tr}$ is the standard training set made of pairs (feature vectors and class labels). The new set $T_u = \{({\bf x_i})\}_{i=1}^{p}$ is made of $p$ unlabelled elements. After a model is built on $T_{\rm tr}$, AL suggests which instances in $T_u$ should be assigned a class label and incorporated into $T_{\rm tr}$ to build a new (refined) model. The task of finding the class label corresponding to an instance in $T_u$ is called \textit{sample query}. AL tries to minimize the cost of querying samples from $T_u$ by focusing on objects with the highest potential to force a change in the current predictive model.  

In our particular study, $T_{\rm tr}$ plays the role of spectroscopic data, where class labels are known. Photometric observations were divided into $T_u$ (objects available for query, whose $r_{mag} \leq 24$) and $T_{\rm te}$ (the remaining non-labelled objects). Final accuracy is measured on test set $T_{\rm te}$. 
The potential distributional discrepancy between spectroscopic and photometric samples is tackled by querying examples from the true target (photometric) distribution  whereas the spectroscopic sample simply serves to provide an initial model amenable to refinement. However, as we showed in sections \ref{sec:TD} and \ref{sec:semi}, the algorithm is effective even when this initial training set is non-informative.

There are many variants of active learning. Some examples include \textit{query synthesis} where the learner can query instances from any region of the input (i.e., feature) space $\mathcal{X}$; or \textit{stream-based selective sampling}, where instances are sampled according to the underlying sample distribution. The most common approach to active learning, and the one adopted in this study, is known as \textit{pool-based learning} where we assume the existence of a dataset $T_u$ from which unlabelled instances can be queried. A key concept in active learning is that each query is associated with a cost; a trade-off then exists between improving the current model by adding more labelled instances, and minimizing the overall cost needed to acquire the corresponding class labels. 
We describe two instantiations of the pool-based sampling approach in the following subsections.

\subsection{Uncertainty Sampling}

One approach to (pool-based) active learning is called \textit{uncertainty sampling}. The main idea is to iteratively train a model $f(\mathbf{x}|\theta)$ on $T_{\rm tr}$ and choose the single instance $\mathbf{x}^*$ from $T_u$ with the highest uncertainty on its class label. After querying the label $y$ of $\mathbf{x}^*$, the algorithm incorporates $(\mathbf{x}^*,y)$ into training set $T_{\rm tr}$, and a new model is induced. This iterative process continues until we reach a user-defined maximum in the number of allowed queries.  

A key component in uncertainty sampling is the criterion to select the instance $\mathbf{x}^*$ with highest uncertainty. Different criteria exist to measure the degree of uncertainty. In every case, it is commonly assumed that the learning algorithm can produce not only a class prediction from model $f(\mathbf{x}|\theta)$, but also a posterior probability $P(y|\mathbf{x},\theta)$, which can be used to quantify the confidence in the prediction. As an illustration, one possible criterion is known as \textit{maximum entropy}; here $\mathbf{x}^*$ is selected based on the class with highest entropy: 

\begin{equation}
\mathbf{x}^* = \argmax_{\mathbf{x}} - \sum_y P(y|\mathbf{x},\theta) \log_2 P(y|\mathbf{x},\theta)
\end{equation}

\noindent
where the right-hand side of the equation computes Shannon's entropy \citep{Cover06}. 

Another popular way to measure uncertainty is known as the \textit{least-confident} approach, where instance $\mathbf{x}^*$ is selected if it minimizes the class posterior probability: 

\begin{equation}
\mathbf{x}^* = \argmin_{\mathbf{x}} P(y|\mathbf{x},\theta).
\end{equation}

One more approach is to select  $\mathbf{x}^*$ by looking for the \textit{minimum margin}, defined as follows: 

\begin{equation}
\mathbf{x}^* = \argmin_{\mathbf{x}} P(y^\prime|\mathbf{x},\theta) - P(y^{\prime\prime}|\mathbf{x},\theta)
\end{equation}

\noindent
where $y^\prime$ and $y^{\prime\prime}$ are the first and second most likely predictions made by $f(\mathbf{x}|\theta)$. 

For our results, we implemented  pool-based least confident AL by using the fraction of votes among the trees when random forest acts as the classifier. The result is an estimation of $P(y^{\prime}|\mathbf{x}, \theta)$, where $y^{\prime}$ is the predicted class, $\mathbf{x}$ are the set of best-fit parameters of equation \ref{eq:bazin} in all 4 filters, and $\theta$ are the parameters of the random forest classifier.

\subsection{Query by Committee}

Within pool-based learning, another popular approach to active learning is known as \textit{Query By Committee} (QBC). The main idea is as follows. Instead of looking for the single instance with the lowest confidence on its prediction, we train a committee or ensemble of different models $\mathcal{C} = \{\theta_i\}$ and look for the single instance where most of the models \textit{disagree} on their prediction. Intuitively, querying such an instance is critical in finding a better location for the final decision boundary. The degree of uncertainty is quantified here in terms of the degree of model disagreement. Specifically, let us refer to every different model according to its parameter set $\theta$. In the previous section we referred to $P(y|\mathbf{x},\theta)$ as the class posterior probability of a model parametrized by $\theta$. Under a committee of different models $\mathcal{C}$, we are now able to compute the average posterior probability of $\mathbf{x}$ (across models in $\mathcal{C}$):

\begin{equation}
\bar{P}_{\mathcal C}(y|\mathbf{x}) = \frac{1}{|\mathcal{C}|} \sum_{\theta} P(y|\mathbf{x},\theta)
\end{equation}

This average probability says something about the uncertainty of the prediction of $\mathbf{x}$, but this time in terms of an ensemble of models. The closer the value to $0.5$\footnote{This is valid when considering binary classifications only.}, the higher the uncertainty of the prediction. We can now look for the single instance that maximizes the entropy of such average probability:

\begin{equation}
\mathbf{x}^* = \argmax_{\mathbf{x}} - \sum_y \bar{P}_{\mathcal{C}}(y|\mathbf{x}) \log \bar{P}_{\mathcal{C}}(y|\mathbf{x})
\label{eq:votee}
\end{equation}

The formula is known as \textit{vote entropy} and is similar to the idea behind uncertainty sampling, but here the class posterior probability is computed through a consensus of committee models. 

In our investigations, we applied three different flavours of QBC. In each case, we used vote entropy as the measure of disagreement. Specifically, let $y$ represent the potential label of a data point $\mathbf{x}$, let $V(y)$ represent the number of votes received by that label, and let $|\mathcal{C}|$ be the number of models in the committee. Our (hard-vote) approximation of eq.~\ref{eq:votee} is defined as follows:

\begin{equation}
\mathbf{x}^* = \argmax_{\mathbf{x}} -\sum_{y}\frac{V(y)}{|\mathcal{C}|}\log_2\frac{V(y)}{|\mathcal{C}|} 
\end{equation}

For the actual software implementation of the different committee models, we used the publicly available \textit{scikit-learn}\footnote{http://scikit-learn.org} Python library, with the default input parameters, unless explicitly specified otherwise.
 
In the following sections, we detail the two QBC flavours that we applied in our experiments.
 
\subsubsection{QBC 1}

In our first implementation of the QBC approach, the committee consisted of a selection of the following $5$ machine learning classifiers:

\begin{itemize}
\item logistic regression,
\item a random forest made up of $100$ trees,
\item a support vector machine classifier,
\item a decision tree,
\item and a k-nearest neighbour classifier with $k=19$.
\end{itemize}

The models used by these classifiers are significantly different from one another, therefore it is reasonable to expect their decision boundaries to differ as well. The rationale behind this approach is that the algorithm would query objects that are not only difficult to classify in a single model, but are challenging across all models, therefore are universally worthy of attention when targeting follow-up observations.





\subsubsection{QBC 2}

In this particular implementation of QBC, we used a committee made up of 15 random forests, each comprising 100 trees. This choice ensures that all committee members have about the same classification performance, as they are indeed different randomized examples of the same model. However, the disadvantage compared to previous selections is that the population of queried objects may be biased by any potential inability of this specific model to fail to generalize well on a given region of the feature space.

\subsubsection{Results}

\begin{figure*}
\includegraphics[width=\textwidth]{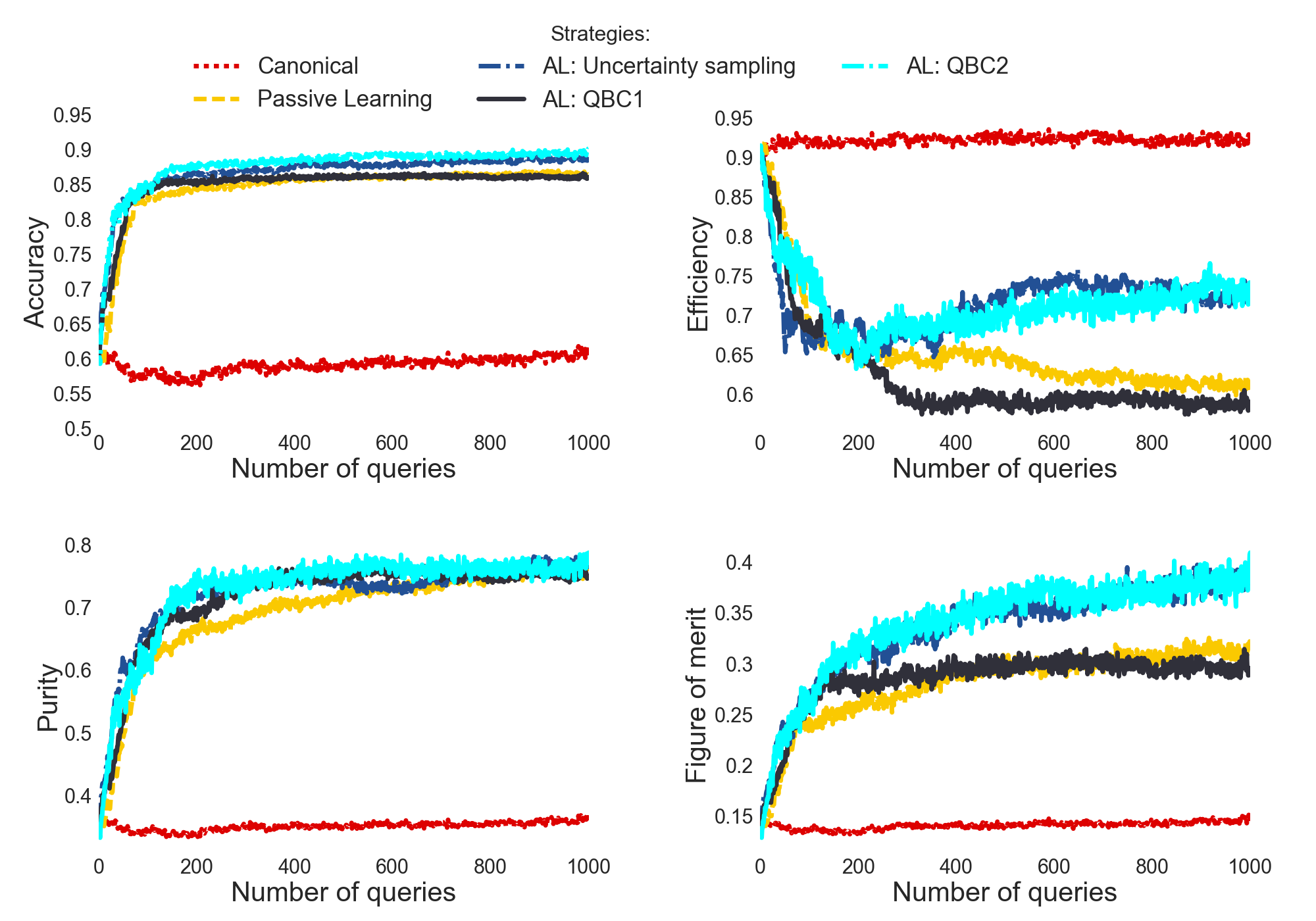}
\caption{Evolution of classification results as a function of the number of queries for the full initial training and full light curve analysis.}
\label{fig:QBC}
\end{figure*}

Figure \ref{fig:QBC} shows how QBC results evolve with the number of queries for the static full training and full light curve scenario. The figure also shows results for the canonical, passive learning, and for AL by uncertainty sampling in order to facilitate comparison (these are the same results shown in figure \ref{fig:fullLC}). We can see that QBC1 results are slightly better than passive learning for very early times - when the impact of non-representativeness is greater. However, the diversity within the committee soon becomes insufficient to tackle the small differences between objects --- and consequently, results are very similar to the passive learning strategy. This can be explained by the fact that, although the elements of the committee are different from each other, they are very simple instances of their class. Increasing the complexity of the elements might provide improved results. 

QBC2 results follow closely those obtained with the uncertainty sampling strategy concerning SNe Ia classification. This was expected, since both approaches utilize a random forest classifier, even though the decision process is somewhat different. In fact, the total number of trees used in each decision is 1500 for QBC2, and only 1000 for uncertainty sampling, which might explain the slight but consistent advantage in accuracy (the total number of correct classifications) for QBC2.
A detailed study of the results of active learning algorithms
focusing on other SN types is outside the scope of this work, but will certainly be the subject of subsequent analysis.

Finally, although the tests we performed were not conclusive regarding the QBC designs, we advise further investigation of this type of algorithm --- especially if new data become available (e.g. PLAsTiCC). This might provide interesting results in the presence of more complex committees and in investigations of other SN types.

\bsp	
\label{lastpage}
\end{document}